\documentclass[twocolumn,amsmath,amssymb,10pt,groupedaddress,a4paper,letterpaper,fleqn]{revtex4}
\usepackage{amssymb}
\usepackage[dvipdfm]{graphicx,color}
\usepackage{epsfig}
\usepackage{dcolumn}
\usepackage{array}
\usepackage{bm}
\usepackage{fancyheadings}
\usepackage{longtable}
\usepackage{multirow}
\usepackage{float}
\usepackage{afterpage}

\setlongtables
\usepackage[breaklinks=true,linkbordercolor={1 1 1}]{hyperref}

\begin{document}
\setcounter{page}{1}

\title{


Neutron Thermal Cross Sections, Westcott Factors, Resonance Integrals, Maxwellian Averaged Cross Sections and Astrophysical Reaction Rates Calculated from  the ENDF/B-VII.1, JEFF-3.1.2, JENDL-4.0, ROSFOND-2010, CENDL-3.1 and EAF-2010  Evaluated Data Libraries
}

\author{ 
B.~Pritychenko,\footnote[1]{Electronic address: pritychenko@bnl.gov}
 S.F.~Mughabghab 
\\
}

\affiliation{ National Nuclear Data Center, Brookhaven National Laboratory, Upton, NY 11973-5000, USA}


\begin{abstract}
 \begin{center}
  \begin{minipage}{6.6in}
   \qquad
   We present calculations of  neutron thermal cross sections, Westcott factors, resonance integrals, 
Maxwellian-averaged cross sections and astrophysical reaction rates for 843 ENDF materials using data from the 
major evaluated nuclear libraries and European activation file. 
Extensive analysis of newly-evaluated neutron reaction cross sections, neutron covariances, 
and improvements in data processing techniques motivated us to calculate nuclear industry and 
neutron physics quantities, produce $s$-process Maxwellian-averaged cross sections and 
astrophysical reaction rates, systematically calculate uncertainties, 
and provide additional insights on  currently available neutron-induced reaction data. 
Nuclear reaction calculations are discussed and new results are presented. 

   \vskip\baselineskip
   \qquad 
  \end{minipage}
 \end{center}
\end{abstract}


\maketitle

\lhead{Neutron Thermal Cross ...}
\chead{BNL REPORT}
\rhead{B. Pritychenko, S.F. Mughabghab}
\lfoot{}
\rfoot{}
\setlength{\headrulewidth}{0.4pt}
\setlength{\footrulewidth}{0.4pt}

\tableofcontents{}

\section{INTRODUCTION}
\label{sec:introduction}

Knowledge of neutron physics  quantities  plays an important role in development of
nuclear energy, national security and nuclear astrophysics 
applications \cite{11Chad,06Mugh,00Bao,05Nak,10Pri}. Their numerical values 
could be extracted from the international collection of Evaluated Nuclear Data File (ENDF) reaction libraries: ENDF/B-VII.1, JEFF-3.1.2, JENDL-4.0, ROSFOND-2010, 
CENDL-3.1  \cite{11Chad,11Kon,11Shi,07Zab,11Zhi} 
and EAF-2010 activation file \cite{10Sub}. To provide user access to the latest 
neutron physics data, we have processed the data files, 
calculated the values and analyzed the results.

In the present work, we consider  neutron elastic  scattering (n,n), fission (n,f) and capture (n,$\gamma$)   
reactions and several parameters that are important for nuclear science and technology applications.
Numerical values of thermal neutron cross sections ($\sigma^{2200}$), Westcott factors ($g_w$), 
resonance integrals (RI),  Maxwellian-averaged cross 
sections  (MACS or $\sigma^{Maxw}$) and astrophysical reaction rates ({\it R(T$_9$)}) were produced in a
systematic approach for Z=1-100 nuclei (materials) using the nuclear 
reaction data, Doppler broadened at multiple temperatures. 

We analyzed the results 
using the {\it Atlas of Neutron Resonances} reference book \cite{06Mugh}, {\it Neutron Cross Section Standards} \cite{09Ca} and KADoNiS database
\cite{06Dil} as benchmarks. This analysis includes over 843 isotopic and elemental neutron
evaluations of importance to the wide range of applications.  The present paper uses 
ENDF/B-VII.1 \cite{11Chad,08Her,09Ob,10Ob,11Ho} and Low-Fidelity neutron cross section covariance data \cite{08Sm,08Lit}, to
provide uncertainty estimates. An extensive discussion on neutron parameter calculations,
potential implications and further recommendations are presented. Since, the current
calculations were completed by March 2012,  only nuclear data libraries
available prior to this date were considered.

Due to a large volume of calculations and size of the nuclear data tables produced in this work, 
the Nuclear Data Sheets  article  contains only the general formalism, discussion of the results  and an 
example of Maxwellian-averaged cross sections table. The extended version of this paper with the complete set of 
nuclear data tables is published in a present report and uploaded to http://arxiv.org/abs/1208.2879 for downloads. The majority of examples and 
figures in NDS paper are given for the recently-released ENDF/B-VII.1 library \cite{11Chad} while the complete sets of  
data for all libraries are published in the Brookhaven report. Finally, we   applied Low-Fidelity covariances to 
ENDF/B-VII.1 neutron physics quantities. The present project organizational chart is shown in Table \ref{Table0}.

\begin{table}
\centering
\caption[Project organizational chart.]
{Project organizational chart. Due to space limitations only MACS at $kT$=30 keV are published in the paper, the complete sets of extracted and calculated data are published in the present report.}\label{Table0} 
\begin{tabular}{|c|c|c|c|}
\hline
{\bf Present Paper} &\multicolumn{3}{c|}{{\bf BNL Report}}\\
\cline{1-4}
(n,$\gamma$) & (n,$\gamma$) & (n,f) & (n,el) \\
\hline\hline
       & $\sigma^{2200}$(293.6) & $\sigma^{2200}$(293.6) & $\sigma^{2200}$(293.6)  \\
       & $\sigma^{2200}$(0)     &                        &                          \\
       & $g_w$       & $g_w$       & $g_w$ \\
       & RI          & RI          & RI    \\
MACS30 &       & MACS30      &  MACS30         \\
       &     & R(T$_9$)    &  R(T$_9$)     \\
       &     & MACS1420    &  MACS1420     \\
       & LowFi       & LowFi       & LowFi \\
\hline
\end{tabular}
\end{table}

\section{EVALUATED NUCLEAR REACTION DATA LIBRARIES}
\label{sec:libraries}
Increasing energy demand, concerns over climate change and dependence on overseas supplies 
of fossil fuels are coinciding to make the strong case for 
 increasing use of nuclear power and provide a strong drive for the 
 nuclear renaissance. Presently, $\sim$20\% of U.S. electrical power is generated by 104 nuclear power plants \cite{10Phy}. 
 These plants provide $\sim$75\% of non-emission generated electricity in the country. Ongoing construction and design of 
 the new power units, 55 worldwide and 6 in the USA, require constant improvements of evaluated nuclear reaction 
 data  that are absolutely essential for reactor operations.

Most evaluated nuclear reaction data files are stored in the databases or libraries using the internationally adopted ENDF-6 format \cite{11He}.  
This format, maintained by the Cross Section Evaluation Working Group (CSEWG) \cite{CSEWG}, provides foundation for the ENDF libraries. 
ENDF is a core nuclear reaction library containing evaluated (recommended) cross sections, 
 neutron spectra, angular distributions, fission product yields, thermal neutron scattering, photo-atomic and other data, with
emphasis on neutron-induced reactions. 

Evaluation and dissemination of nuclear reaction data are coordinated by the CSEWG 
and the U.S. Nuclear Data Program (USNDP) \cite{CSEWG,USNDP,06Pri} in the 
USA and by the Working Party on International Nuclear Data Evaluation Co-operation (WPEC) \cite{WPEC} worldwide.
 
Broad international effort led to the development of a variety of ENDF-6 formatted libraries: ENDF/B in the USA \cite{06Chad,01Cse}, JEFF in Europe
\cite{11Kon}, JENDL in Japan \cite{02Shi}, ROSFOND and BROND in Russia \cite{07Zab,BROND} and CENDL in China \cite{11Zhi}. USA,
European Union, Japan, China and Russia are continuing to heavily invest into evaluated nuclear reaction libraries research and development. This effort produced
successful releases of  ENDF/B-VII.1 and ENDF/B-VII.0 (U.S. 2011 and 2006), JEFF-3.1.2 (Europe 2012), JENDL-4.0 (Japan 2010), ROSFOND (Russia 2008-2010), CENDL-3.1 (China 2009)  libraries and EAF-2010 (Europe 2010) activation file. 
Most of these libraries are based on the original ENDF evaluations, while ROSFOND ({\it ROSsijsky File Otsenennykh Neutronnykh Dannykh}, in Russian) is completely based on pre-selected and often adjusted evaluations from ENDF/B-VII.0, ENDF/B-VII.beta2, ENDF/B-VI.8, JEFF-3.1, JENDL-3.3, CENDL, BROND-3, FOND-2.2 and 
European Activation File (EAF2003) \cite{06Chad,01Cse,11Kon,02Shi,07Zab,BROND,07Eaf}. The ROSFOND library selection process 
has produced a unique blend of 686 high-quality nuclear industry evaluations that spans over $\sim$98$\%$ of {\it s}-process nucleosynthesis nuclei.

\subsection{Evaluated Neutron Cross Section Data}
 
Frequently, ENDF cross sections  are represented in groupwise (averaged over broad energy interval) and pointwise (energy-dependent or ``continuous") formats \cite{10Mac,yu09}. 
The first representation is often used in reactor physics calculations, while the second one is better
suited for nuclear physics applications. In this paper, we will consider several uses of evaluated data for slow and fast neutrons 
and compare obtained values with EXFOR database \cite{EXFOR}, 
{\it Atlas of Neutron Resonances} reference book \cite{06Mugh} and other commonly-used standards and benchmarks \cite{09Ca,06Dil}. 
 ENDF library evaluations do not contain a complete set of cross sections in a single file. Codes such as PREPRO \cite{07Cul} and NJOY \cite{10Mac} 
are used for processing the files to produce Doppler broadened neutron cross sections   
within the ENDF range of energies from 10$^{-5}$ eV to 20 MeV.

In the present work, ENDF data were preprocessed with 0.1$\%$ precision within T=0 - 1000$^{\circ}$ K temperature range using code PREPRO \cite{07Cul}. 
Zero degree temperature is of interest to {\it neutron cross section standards} \cite{09Ca},
of space exploration and cryogenic detector development \cite{00Abu}. Many neutron cross section values were measured  
at thermal energies,  defined as 0.0253 eV for 2200 m/s neutrons. 
The famous plasma physics ratio 1 eV = 11605$^{\circ}$ K  leads to the second processing temperature of 293.6$^{\circ}$ K.  
To satisfy the needs of nuclear reactor operators, data were also processed at 575$^{\circ}$ K \cite{06West,10Can}.

Data analysis for several processing temperatures provided an additional quality assurance for present calculations. 
Elastic neutron scattering cross sections strongly depend on the temperature. ENDF/B-VII.1 reaction cross sections 
for $^{1}$H(n,n) at T=0 and 293.6$^{\circ}$ K are shown in the  Fig.\ref{fig0} and their values are listed in Table II.
\setcounter{figure}{0}
\begin{figure}
\begin{center}
\includegraphics[height=7cm]{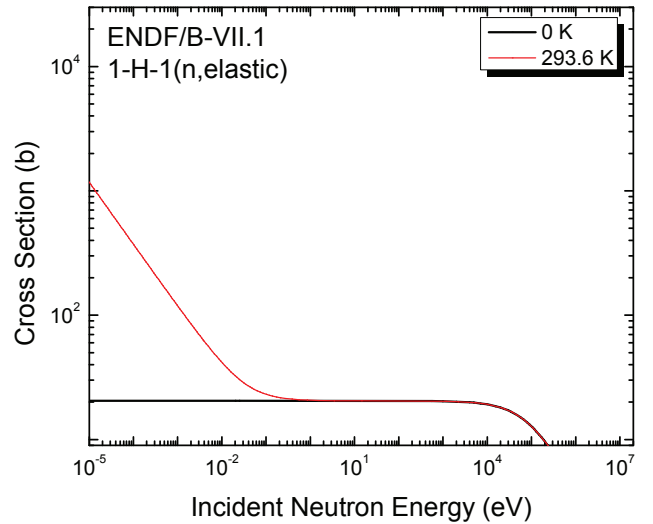}
\caption{ENDF/B-VII.1 evaluated elastic neutron  cross sections for $^{1}$H at 0 and 293.6 K are shown in black and red, respectively.}
\label{fig0}
\end{center}
\end{figure}
Temperature selection affects thermal neutron cross sections, Westcott factors and to a lesser extent resonance integrals. 
Resonance integral values are influenced by temperatures $>$ 5000$^{\circ}$ K or in rare case of strong resonances near thermal energy, 
i.e $^{237}Np(n,\gamma)$ \cite{06Mugh}. Maxwellian-averaged cross sections for $s$-process nucleosynthesis and $^{252}$Cf spectra are 
almost identical within the  processing temperature range. Due to space limitations neutron elastic thermal cross sections  
are given only at T=0 and 293.6$^{\circ}$, the rest of the quantities are  listed at T=293.6$^{\circ}$.

\section{CALCULATION OF NEUTRON THERMAL CROSS SECTIONS, WESTCOTT FACTORS, RESONANCE INTEGRALS,  MAXWELLIAN-AVERAGED CROSS SECTIONS AND ASTROPHYSICAL REACTION RATES}
\label{sec:calculation}
We performed calculations of neutron thermal cross sections, Westcott factors, resonance integrals,  
Maxwellian-averaged cross sections and reaction rates 
using the evaluated neutron library data \cite{11Chad,11Kon,11Shi,07Zab,11Zhi,10Sub}. The definitions of neutron physics quantities  
 are presented below.

Since neutron thermal cross sections in the lab reference system are tabulated in Doppler-broadened ENDF evaluations, we simply extracted these values from the evaluations.

Westcott g-factor, $g_w$, is the ratio of Maxwellian-averaged cross section to the 2200 m/s (thermal) cross section \cite{55We} 
\begin{equation}
\label{myeq.west1}
g_w = \frac{\sigma^{Maxw}}{\sigma^{2200}}.
\end{equation}
The $g_{w}$-factor is temperature dependent  \cite{55We,inter,99Ho,07Ch} and its value is close to 1 for most nuclei where $\sigma (n,\gamma) \sim \frac{1}{\upsilon}$. 

The epicadmium dilute resonance integral (RI) \cite{06Mugh} for a particular reaction $\sigma_R (E)$ in $1/E$ spectrum is expressed by
\begin{equation}
\label{myeq.res1}
RI =  \int_{E_c}^{\infty} \sigma_R (E) \frac{dE}{E},
\end{equation}
where $E_c$ is determined by cadmium cutoff energy ($E_c$=0.5 eV). The cutoff energy depends on the Cd shield thickness, 
its values are approximately 0.5 and 0.55 eV for $\sim$1 and $\sim$1.5 mm cadmium  shields in case of beam geometry \cite{68De}, respectively. 
 The absorbing nuclide is assumed to be present in such small or diluted quantities that there is no perturbation of the neutron slowing down spectrum  \cite{02Ba}. 
This is why, the resonance integral is called infinitely dilute RI.

Average cross sections for Maxwellian spectrum  are defined as 
\begin{equation}
\label{myeq.max1}
\sigma^{Maxw}(kT) =  \frac{\langle \sigma \upsilon \rangle}{\upsilon_T},
\end{equation}
where $\upsilon$ is the relative velocity of neutrons and a target nuclide and $\upsilon_{T}$ is the mean thermal velocity given by 

\begin{equation}
\label{myeq.max2}
\upsilon_{T} = \sqrt{\frac{2kT}{\mu}},
\end{equation}
where $\mu$ is the reduced mass. 

Maxwellian-averaged cross sections (MACS) can be expressed as \cite{10Pri}   

\begin{equation}
\label{myeq.max3}
\sigma^{Maxw}(kT) = \frac{2}{\sqrt{\pi}} \frac{a^{2}}{(kT)^{2}}  \int_{0}^{\infty} \sigma(E^{L}_{n})E^{L}_{n} e^{- \frac{aE^{L}_{n}}{kT}} dE^{L}_{n},
\end{equation}
where $a = m_2/(m_1 + m_2)$, {\it k} and {\it T} are the Boltzmann constant and temperature of the system, respectively. $E^{L}_{n}$ is neutron energy in the laboratory system and $m_{1}$ and $m_{2}$ are masses of 
the neutron and the target nucleus, respectively.

The astrophysical reaction rate, $R$, is defined as $R$ = $N_{A}$$\langle \sigma \upsilon \rangle$, where $N_{A}$ is the Avogadro number. To express reaction rates in [$cm^{3}$/mole s] units, an additional factor of $10^{-24}$ is introduced; $\upsilon_{T}$ is in units of [cm/s] and temperature, $kT$, in units of energy ({\it e.g.} MeV) is related to that in Kelvin ({\it e.g.} 10$^{9}$ K) as $T_{9}$=11.6045$kT$.
\begin{equation}
\label{myeq.rrates}
R(T_{9}) = 10^{-24}N_{A}\sigma^{Maxw}(kT)\upsilon_{T}.
\end{equation}

These equations were used to  calculate integral values using the original evaluated neutron data 
from the ENDF libraries within the typical range of energies.

\subsection{Calculation of ENDF Integral Values}

Previously published computations of Maxwellian-averaged cross sections and astrophysical reaction rates  \cite{10Pri} were based on the Simpson integration method for the linearized ENDF cross sections (MF=3). 
This method provided quality integral values. However, the degree of precision was within $\sim 1 \%$ \cite{05Nak,10Pri}. 
This limitation can be overcome in the linearized ENDF files because cross section value is linearly-dependent on energy within a particular bin \cite{10BPri}
\begin{equation}
\label{myeq.int1}
\sigma (E) = \sigma_{1} + (E-E_1)\frac{\sigma (E_2) - \sigma (E_1)}{E_2 - E_1}, 
\end{equation} 
where $\sigma (E_1),  E_1$ and  $\sigma (E_2),  E_2$ are pointwise cross section and energy values for the energy bin.

In the present work, we  calculate definite integrals applying the Wolfram Mathematica online integrator \cite{09Math}. 
Summing definite integrals for all energy bins and sufficiently dense grid produces an exact integral value for parameter of interest.

\begin{figure}
\begin{center}
\includegraphics[width=0.95\columnwidth]{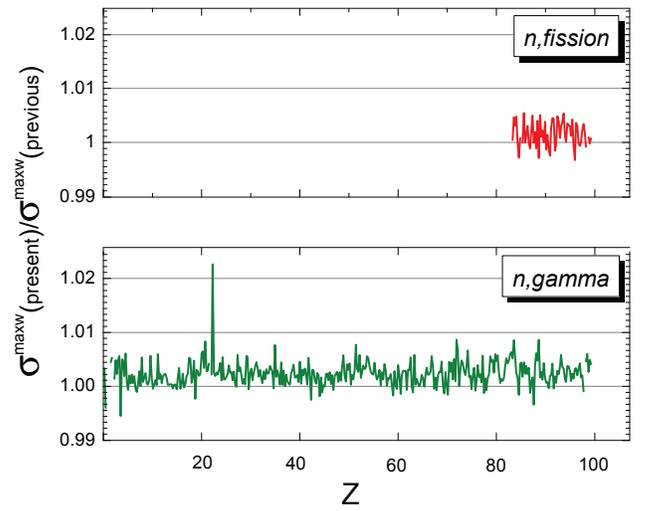}
\caption{Calculated ENDF/B-VII.0 library MACS ratios for neutron-induced fission and capture reactions using the present and previous \cite{10Pri} methods.  Horizontal scale is relative.}
\label{fig2}
\end{center}
\end{figure}

To validate this approach ENDF/B-VII.0 Maxwellian-averaged cross sections  were re-calculated and compared with our previously published Simpson integration values \cite{10Pri}. 
Fig. \ref{fig2} shows a good agreement between two calculations with the exception of $^{79}$Se where, due to lack of experimental data, ENDF/B-VII.0 
 evaluated capture cross sections have problem with merging of low- and high-energy data.

\subsection{Calculation of Cross Section Uncertainties}

\begin{figure}
\begin{center}
\includegraphics[height=7cm]{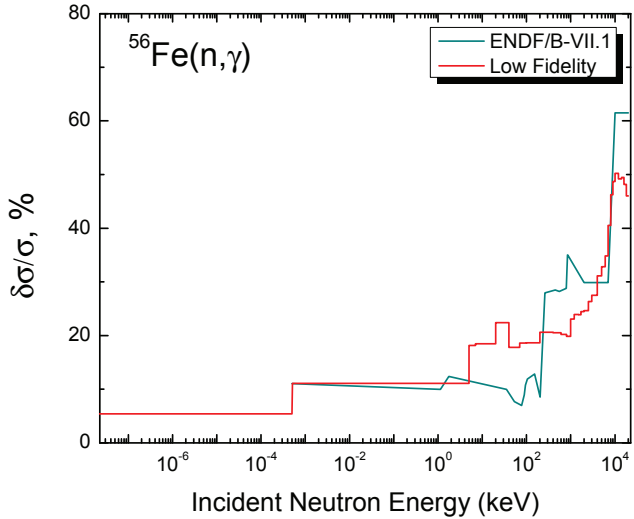}
\caption{ENDF/B-VII.1 and Low-Fidelity groupwise cross section uncertainties for $^{56}$Fe(n,$\gamma$) reaction are shown in green and red colors, respectively.}
\label{fig1}
\end{center}
\end{figure}

Major evaluated neutron libraries contain limited sets of covariance files. In this work, we will consider 190 ENDF/B-VII.1 covariance files mostly coming from structural, fission and actinide materials 
and 387 covariance files from the Low-Fidelity project \cite{11Chad,08Sm,08Lit}. 
The Low-Fidelity   covariance data practically cover the whole ENDF/B-VII.0 range of nuclei. This project was not an official part of  
ENDF/B-VII.0 release. However, the selection of evaluations was motivated by ENDF/B-VII.0 material list. 
An example of the $^{56}$Fe(n,$\gamma$) groupwise cross section uncertainties for ENDF/B-VII.1 library and the Low-Fidelity projects is shown in Fig. \ref{fig1}. 
This Figure shows general consistency between two cross section uncertainties within the ENDF range of energies.

As in the ENDF/B-VII.1 reference paper \cite{11Chad} we calculated  Maxwellian-averaged cross section and reaction rate uncertainties.  
For comparative purposes, the  Low-Fidelity covariances were used to produce uncertainties for ENDF/B-VII.1 integral values, as shown in the Report Tables XVII-XIX.

\section{RESULTS AND DISCUSSION}
\label{sec:results}
In this section we present computations of neutron physics quantities and discuss their implications. Neutron thermal cross sections were extracted from the ENDF libraries while  Westcott factors, resonance integrals,  
Maxwellian-averaged cross sections and reaction rates for the ENDF/B-VII.1, JEFF-3.1.2, 
JENDL-4.0, ROSFOND-2010, CENDL-3.1 libraries and EAF-2010 activation file have been calculated. Present values are in agreement 
with our earlier results for the Sigma Web Interface \cite{10Sig,08Sig} at T=0 and 300$^{\circ}$ K. 
An extensive analysis of the ENDF astrophysical data has been  performed in our previous work \cite{10Pri}. 
In this work, we will extend the analysis to neutron thermal cross sections, Westcott factors and resonance integrals.

\subsection{Neutron Thermal Cross Sections}
Neutron thermal cross section values  for  elastic scattering, fission and capture are shown in the current report. 
These values are compared with {\it Atlas of Neutron Resonances} and {\it Neutron Cross Section Standards} evaluations \cite{06Mugh,09Ca}.

\begin{figure}
\includegraphics[width=0.95\columnwidth]{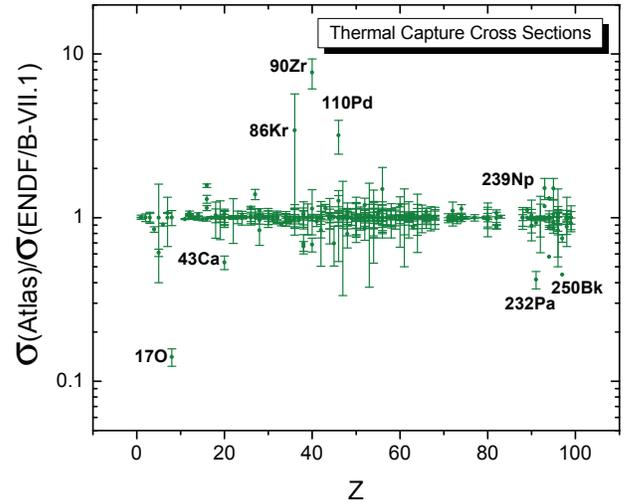}
\caption{Ratio of thermal neutron capture cross sections for {\it Atlas of Neutron Resonances} to ENDF/B-VII.1.}
\label{IQ.1}
\end{figure}

\begin{figure}
\includegraphics[width=0.95\columnwidth]{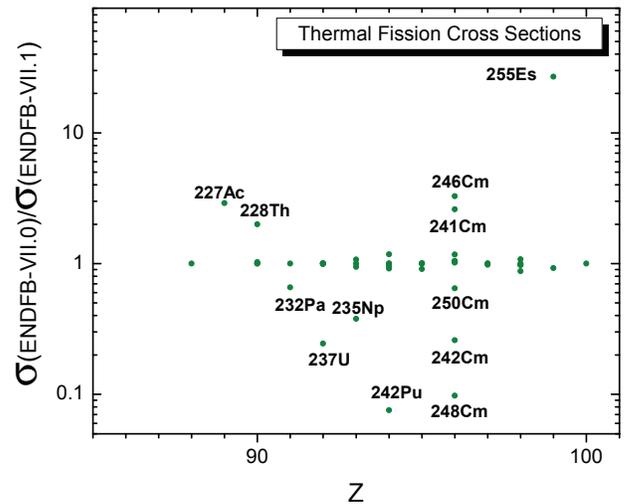}
\caption{Ratio of  thermal neutron fission cross sections for ENDF/B-VII.0 to ENDF/B-VII.1. Where discrepancies are evident, VII.1 values are thought to be more accurate.}
\label{IQ.2}
\end{figure}

ENDF/B-VII.1 thermal cross section ratios for neutron capture and fission are shown in
Figs. \ref{IQ.1}, \ref{IQ.2} and Tables III-IV of the report.  Using  visual
inspection we notice deviations for light and medium nuclei and
minor actinides evaluations. These differences, in the low- and 
medium-Z region, are attributed to the lack or insufficient
experimental data for $^{17}$O, $^{43}$Ca, $^{86}$Kr, $^{110}$Pd, and 
the recent re-evaluation of $^{90,91}$Zr.
In the minor actinide region, deviations are actually in the JENDL-4.0 evaluations \cite{11Shi}, 
since these were adopted by ENDF/B-VII.1 library.

In this subsection, we briefly discuss the discrepancies observed
between the ENDF/B-VII.1  and the {\it Atlas of Neutron Resonances} \cite{06Mugh} regarding the thermal cross sections.

\begin{itemize}
\item  $^{17}$O

    The  ENDF/B-VII.1 thermal capture cross section, 3.8 mb, for  $^{17}$O is
 attributed to a misquoted direct capture cross section  calculated by
 Mughabghab \cite{81Mu}. The actual correct value is 0.38 mb. The {\it Atlas} value is based
 on measurements reporting a value of 0.538$\pm$0.065 mb.
 Since the thermal cross section contributes about half of its value to the capture resonance
 integral, a similar discrepancy would result
 also in this quantity as shown in Fig. \ref{IQ.3}.

\item  $^{43}$Ca

 The ENDF/B-VII.1 file for  $^{43}$Ca is taken from the Petten evaluation.
 In its documentation, it is stated that the thermal capture cross
 section, $\sigma$(n,$\gamma$)= 6.2 b, is adopted from \cite{81Mu}. In order to reproduce this
 value, a negative resonance was added in the Petten evaluation. However, the calculation of
 the thermal cross section from the resonance parameters yields 11.66 b. Note that, according to  the {\it Atlas},
 the bound level contributes only 1.61 b to the thermal capture cross
 section.

\item $^{86}$Kr

      The ENDF/B-VII.1 file for $^{86}$Kr is a product of the WPEC
international cooperation. In its documentation, it is stated 
``a negative resonance was added at -20 keV so as to reproduce the thermal
capture cross section of 3 mb given by \cite{81Mu}." This evaluated file 
does not contain a negative energy resonance and the table of thermal
cross sections in the documentation reports a calculated thermal
capture cross section of 0.877 mb  in agreement with  the {\it Atlas} calculation.
 Note that since the recommended thermal capture
cross section in the {\it Atlas} is 3$\pm$2 mb, there is no need to invoke
a negative energy resonance.

\item $^{90,91}$Zr

For $^{90}$Zr, the  ENDF/B-VII.0 thermal capture cross section, 0.077 b, was adopted from the {\it Atlas} recommendations. This value was 
derived using the subtraction method by considering a thermal capture cross section of 0.830$\pm$0.082 b for $^{91}$Zr \cite{81Lo}. However, a more 
recent measurement reported a low limit of 1.30$\pm$0.04 b for the $^{91}$Zr thermal capture cross section \cite{07Na} showing that the previously-derived 
value for $^{90}$Zr was overestimated. On this account, the parameters of the bound level for $^{90}$Zr in ENDF/B-VII.1 evaluation were deleted. 
With this change, the calculated 2200 m/sec thermal capture cross section for $^{90}$Zr becomes 9.97 mb, which is now in good agreement with Lone's measurement, 14$^{+8}_{-4}$ mb \cite{81Lo}.

\item $^{110}$Pd

   The ENDF/B-VII.1  evaluated file for $^{110}$Pd gives  a thermal
   capture cross section of 0.229 b, whereas the {\it Atlas} evaluation reports
   a   corresponding capture cross section more than twice this value, i.e.
   0.73$\pm$0.17 b. The {\it Atlas} value is based on two isomeric ratio
   measurements by Namboodiri   {\it et al.} \cite{Na66} and Dzantiev {\it et al.} \cite{Dz57}
    which are in
   excellent agreement, while the ENDF/B-VII.1 value is based on an older
   evaluation \cite{81Mu}.

\item $^{238}$U

There is a disagreement between {\it Atlas of Neutron Resonances} \cite{06Mugh} and ENDF/B-VII.1 thermal fission cross sections for $^{238}$U.  In this case, fission threshold is well above 
thermal energy, but sub-threshold thermal neutron fission takes place. 
Experimental measurement of D'Hondt {\it et al.} \cite{84Dh} indicates lower than  ENDF/B-VII.1 library calculated value and explains the disagreement.

\item $^{239}$Np 

  The ENDF/B-VII.1 file for $^{239}$Np is adopted from the JENDL-4.0 evaluation,
  which adopted a thermal capture cross section of 45.02 b. This cross section is
  attributed to Stoughton and Halperin \cite{St59}, who reported a thermal
  reactor value  for this nuclide of 80$\pm$20 b. To convert this value
  to a thermal cross section, the  ratio of the epithermal flux to the thermal
  flux, as well as the capture resonance integral, must be known. Since neither of
  these quantities are determined, nonetheless, these authors made some
  estimates and arrived at a thermal capture cross section of 45$\pm$15 b.
    On the other hand, Mughabghab \cite{06Mugh}
   arrived at a thermal reactor capture cross section of 68$\pm$10 b,
   based on activation measurements. Even though the two reported capture 
   cross sections are consistent within their uncertainty limits, nevertheless
   it is interesting  to explore the comparative method \cite{Mu12} to determine its
   applicability in this mass region for other nuclei, such as  illustrated below for $^{250}$Bk.
     Adopting a 2200 m/s capture cross section
   of 175.9$\pm$2.9 b for   $^{237}$Np and a nuclear level density  parameter of a$_{ld}$=26 MeV$^{-1}$ for  $^{239}$Np \cite{06Mugh},  63 b is readily calculated for  the 2200 m/s
   capture  cross section for  $^{239}$Np. This value is in agreement with the {\it Atlas} evaluation. 

\item $^{250}$Bk

  The ENDF/B-VII.1 file for $^{250}$Bk is also adopted from  the JENDL-4.0 evaluation,
  which estimated a thermal capture cross section of 780.5 b
  based on a computer code calculation  \cite{Io07}. The corresponding value in
  the {\it Atlas}, $\sim$350 b, is based on a rough estimate reported in \cite{Bo70}.
  A reasonable estimate was made by one of the authors (SFM) using the 
  comparative method by adopting a thermal capture cross section for $^{249}$Bk
  as 746$\pm$40 b and a nuclear level density parameter a$_{ld}$=29 MeV$^{-1}$ for $^{250}$Bk \cite{06Mugh}.
   The result for the 2200 m/s capture
  cross section is  666$\pm$70 b which is consistent, within 17$\%$ uncertainty, with the JENDL-4.0 estimate
  but not that of Bol'shov {\it et al.} \cite{Bo70}.
\end{itemize}

\subsection{Westcott Factors}

\begin{figure}
\includegraphics[width=0.9\columnwidth]{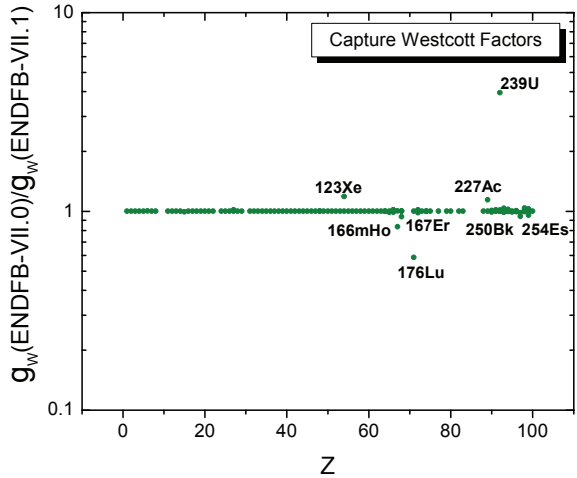}
\caption{Ratio of thermal neutron capture Westcott factors for ENDF/B-VII.0 to ENDF/B-VII.1 libraries.}
\label{IQ.6}
\end{figure}

\begin{figure}
\includegraphics[width=0.9\columnwidth]{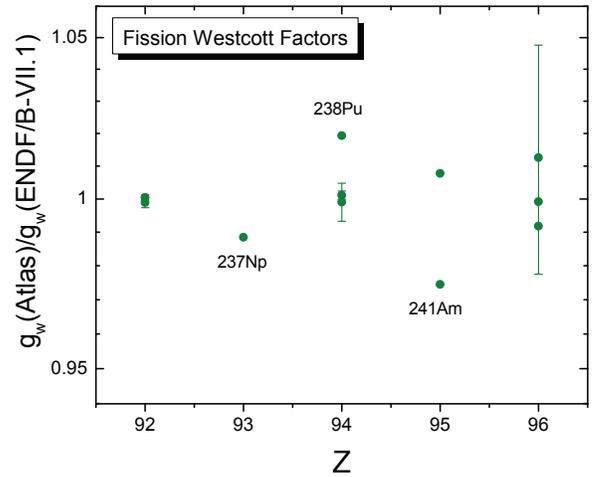}
\caption{Ratio of thermal neutron fission Westcott factors for Atlas to ENDF/B-VII.1 library.}
\label{IQ.WF}
\end{figure}

Calculated Westcott factors are given in the Tables V-VII of this report. Fig. \ref{IQ.6} shows the ratio of capture Westcott factors 
between ENDF/B-VII.1 and ENDF/B-VII.0 libraries, and indicates large changes for $^{239}$U and $^{176}$Lu in neutron capture cross section evaluations.  
These  deviations reflect the recent improvements of ENDF/B-VII library evaluations \cite{11Chad}, 
where Westcott factors evolved from 3.997 to 0.989 and 
from 1.002 to 1.713 for $^{239}$U and $^{176}$Lu, respectively. The last number agrees well with the recommended value of 1.75 \cite{06Mugh}. 
Smaller deviations as in unstable $^{123}$Xe nucleus are due to adoption of 
new evaluations in ENDF/B-VII.1 library and lack of experimental data for this material. {\it Atlas of Neutron Resonances} references book \cite{06Mugh} includes compilation of experimental fission g$_{w}$ factors. 
Fig. \ref{IQ.WF} shows a ratio between the Atlas and calculated ENDF/B-VII.1 fission Westcott factors.

Complete calculation of capture and fission g$_{w}$ factors reveals that most of them are close to 1 with an exception 
of  non-$1/\upsilon ~\sigma (n,\gamma)$ nuclei: $^{113}$Cd, $^{135}$Xe, $^{149}$Sm,  $^{151}$Eu, $^{176}$Lu, $^{182}$Ta, $^{239}$Pu,  
$^{249}$Bk \cite{06Mugh}.  Strong resonances in the thermal energy region, such as 0.29562 eV resonance in $^{239}$Pu \cite{06Mugh}, 
are often responsible for Westcott factor temperature variations. Current calculation provides a very powerful tool for finding the deficiencies in evaluated neutron libraries. 
There are notable Westcott factor issues for unstable isotopes of   $^{151}$Gd, 
$^{155,157,158}$Tb, $^{169,172}$Lu, $^{182}$Re and $^{232}$Pa in ROSFOND 2010 library evaluations where 
neutron capture cross section values suddenly increase at $\sim$ 0.1 eV  due to step function problems.

\begin{figure}
\includegraphics[width=0.9\columnwidth]{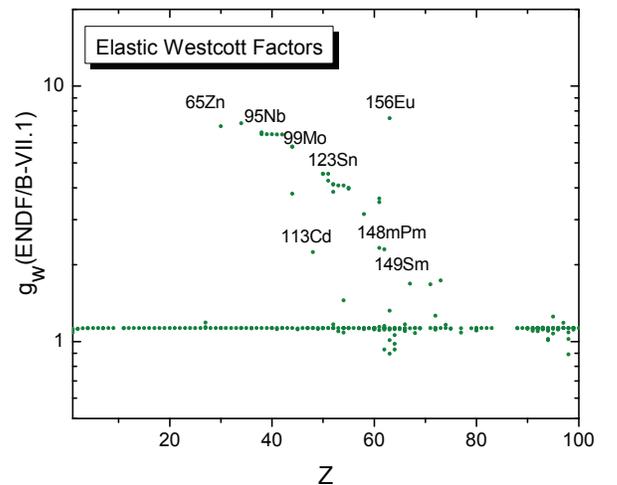}
\caption{ENDF/B-VII.1 library thermal neutron elastic Westcott factors.}
\label{IQ.WE}
\end{figure}

The analysis of elastic scattering g$_{w}$ factors is more difficult due to  lack of experimental data. The majority of shown in Fig. \ref{IQ.WE}
calculated ENDF/B-VII.1 factors are close to 1.1. This value is consistent with the previous calculation by the JEFF team \cite{06Ko}. 
Further analysis for  elastic scattering Westcott factors indicates 29 ENDF/B-VII.1 values for mostly unstable materials, where factors are within 2-8 range. In cases of $^{113}$Cd, $^{148m}$Pm and  $^{149}$Sm evaluations, 
where the Westcott factor is within 2-3 range, the  deviations from unity could be explained by neutron spectra shapes and resonance structures. 
For the rest of 26 cases ($^{65}$Zn,$^{99}$Mo,$^{95}$Nb,$^{123}$Sn,$^{156}$Eu,...), where   Westcott factor value is 
within 3-8 range, the strong deviations are due to improper merging of evaluated cross section curves that results in step-function structures.

\subsection{Resonance Integrals}

The resonance integrals for scattering, fission and capture are calculated 
with the aid of  Eq. \ref{myeq.res1} with an upper energy  limit of 20 MeV.The results are
summarized in Tables VIII - X of the BNL report. For capture reactions,
the dominant contribution to this integral  comes from the energy region below
a few keV. In contrast, for threshold reactions and subthreshold fission, this is
not the case, since the major contribution comes from the energy region above the threshold energy.  
For scattering integrals the data analysis is hampered by lack of experimental data.

\begin{figure}
\includegraphics[width=0.95\columnwidth]{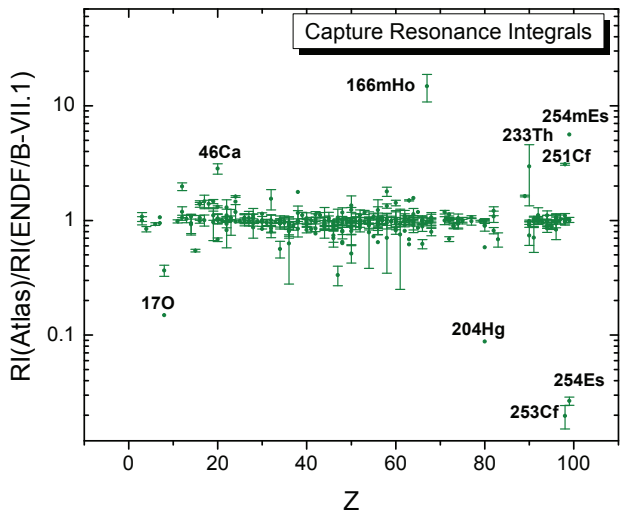}
\caption{Ratio of neutron capture resonance integrals for {\it Atlas of Neutron Resonances} \cite{06Mugh} to ENDF/B-VII.1.}
\label{IQ.3}
\end{figure}

\begin{figure}
\includegraphics[width=0.95\columnwidth]{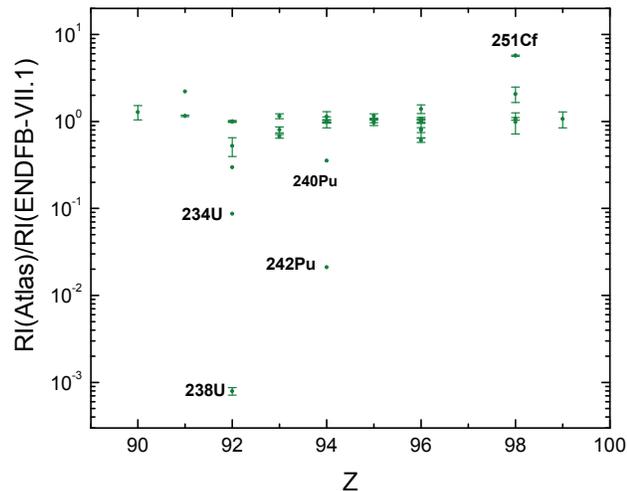}
\caption{Ratio of neutron fission resonance integrals for Atlas to ENDF/B-VII.1. Where discrepancies are evident, Atlas values are thought to be more accurate.}
\label{IQ.4}
\end{figure}

The ratio of the Atlas resonance integrals to those of ENDF/B-VII.1 for capture and fission
are shown in Fig. \ref{IQ.3} and \ref{IQ.4}, respectively. Most of the values are consistent with available measurements \cite{06Mugh,06Lei,09Trb}. 
The observed discrepancies in these quantities are discussed in some detail below.

\subsubsection{Neutron Capture Resonance Integrals}

\begin{itemize}
\item  $^{17}$O

 The large  discrepancy in the ENDF/B-VII.1  is previously explained
 in terms of the incorrect capture cross section for this nuclide.

\item $^{46}$Ca

The ENDF/B-VII.1 file was taken from the JEFF-3.1 evaluation, which does not contain resonance parameters. This nuclide 
with a very low natural abundance of only 0.004$\%$ has a strong s-wave resonance at 28.4$\pm$1.6 keV \cite{06Mugh}. 
Its contribution to capture integral is 1.16 b which is understandably at variance with the ENDF/B-VII.1   
capture cross section. As a result, the capture file of this evaluation has to be updated to take into consideration 
the effect of this strong s-wave resonance to the capture integral as well as the 30-keV Maxwellian-averaged capture cross section.

\item  $^{166m}$Ho

There is a large discrepancy in the ENDF/B-VII.1 calculated capture resonance 
integral for this nuclide, 0.676 kb (based on the resonance parameters \cite{Ma92})
and the measured  value \cite{Ka02},  10.0$\pm$2.7 kb. This is largely due to the fact that the 
authors \cite{Ka02} assumed the wrong spin statistical factor for $^{166m}$Ho. The evaluated resonance parameters for
this nuclide have to be modified in the future.

\item  $^{204}$Hg

The ENDF/B-VII.1 file is based on JENDL-3.3 evaluation, where no resonance parameter information is
given; only pointwise cross section is presented. The thermal capture cross section, 0.43 b, is in agreement with the
{\it Atlas} value \cite{06Mugh} but not the capture resonance integral which gives a value of 2.72 b. In
contrast, the {\it Atlas} value, 0.80$\pm$0.04 b, is based on measurements.

 \item  $^{233}$Th
  
  The capture resonance integral of $^{233}$Th, which was adopted from JENDL-4.0, is 574 b, while the measured
  value in the {\it Atlas} is  1700$\pm$900 b. Since the ENDF/B-VII.1 2200 m/sec capture cross section   
 is 1290.6 b, this  shows that there is no contribution to the capture integral from the resonance region, which
 is in disagreement with the measurement, giving a reduced resonance integral of 1100$\pm$930 b \cite{Ch04}.

  \item  $^{251}$Cf
 
 The resonance capture integral for the ENDF/B-VII.1 file for $^{251}$Cf, adopted from JENDL-4.0 is 519 b for a 
 Cd cutoff of 0.5 eV whereas the  corresponding {\it Atlas} value is 1600$\pm$30 b; the latter value is based on measurements \cite{Ha71}.
  The  large discrepancy can  be attributed to  positive energy at  0.389$\pm$0.001 eV, which
 is close to the cadmium cutoff energy of 0.5 eV. To illustrate the point, if the Cd thickness used in the measurements
  is such that it corresponds to an energy cutoff of 0.40 eV, then the calculated capture resonance integral would be 2038 b instead of 519 b.

 \item  $^{253}$Cf
 
  Measurements of the capture resonance integral and resonance parameters are not yet available for $^{253}$Cf.
   Estimates of the capture cross
  section and capture resonance integral, $\sigma_{\gamma}$= 12.0 and $I_{\gamma}$=12.0 b, were made by Benjamin {\it et al.} \cite{Be75}
  from  multigroup cross section considerations. The ENDF/B-VII.1 value, based on the JENDL-4.0 evaluation,  is
  658 b and is obtained by the CCONE computer code \cite{Io07}. On the other hand the {\it Atlas} recommendation, 13$\pm$3 b 
   is  largely based on Benjamin {\it et al.} estimate \cite{Be75}.

 \item  $^{254g}$Es

  The {\it Atlas} resonance integral for capture, 18.2$\pm$1.5 b, is based on the measurements of Halperin {\it et al.} \cite{Ha85}
  while the ENDF/B-VII.1 (JENDL-4.0) value, $I_{\gamma}$= 683 b, is probably based on CCONE code \cite{Io07}. 

\item  $^{254m}$Es

 Capture cross section measurements and resonance information are not available for this nuclide.
 The ENDF/B (JENDL-4.0)  assumed a capture cross section of 250 b at 0.0253 eV. The ratio of fission to capture cross
 section was then calculated by CCONE code and the capture cross section above 1 eV was deduced. From this capture cross section
 and the thermal value, a capture resonance integral of 178 b was obtained. On the other hand  an  approximate value of
 1000 b, as reported in the {\it Atlas}, is based on measurements.  
\end{itemize}

\subsubsection{Fission Resonance Integrals}

It is important to note that for  $^{234}$U, $^{238}$U, $^{237}$Np, $^{240}$U,  $^{242}$Pu target nuclei, where
subthreshold fission was observed, the calculated fission integrals,  I$_{f}^{c}$ reported in
the Atlas of Neutron Resonances \cite{06Mugh}  correspond to subthreshold fission values.   
Since these calculations were carried out with the recommended resonance parameters \cite{06Mugh} 
and did not include contributions from the unresolved resonance and the fast energy regions. 
These quantities then  represent subthreshold fission integrals and are a small fraction of the total fission integral. 
To illustrate the point, the subthreshold fission integral of $^{237}Np$ is 0.650 b 
calculated from the resonance parameters \cite{06Mugh} while the fission integrals with  upper energy
limits of 10 MeV and 20 MeV are 5.43 b and 6.95 b, respectively, on the basis of the ENDF/B-VII.1 evaluation. 
It is of interest that the only reported measured value for this nucleus gives a value of 4.70$\pm$0.23 b  \cite{Ko86}. 

The discrepancies noted in Fig. \ref{IQ.4}  for $^{234}$U, $^{238}$U,$^{240}$U, $^{242}$Pu 
are explained in terms of the foregoing discussion. The discrepancy for $^{251}$Cf capture integral is noted in the previous section. The
same situation occurs for the fission integral. The Atlas value for the fission integral
is based on measurements by the activation method while the ENDF/B-VII.1 is calculated 
with the resonance parameters of   $^{251}$Cf which are not well determined because of lack of highly-enriched samples for this isotope \cite{An91}.

\subsection{Maxwellian-averaged Cross Sections}
Maxwellian-averaged cross sections have been calculated for two temperatures ($kT$) of  30 keV and 1420 keV. The first temperature is the commonly accepted value  
in stellar nucleosynthesis $s$-process compilations, while the second one closely reproduces a $^{252}$Cf neutron fission spectrum. 

The slow-neutron capture (s-process) is responsible for creation of $\sim$50 $\%$ of the elements
beyond iron. In this region, neutron capture becomes dominant because of the increasing
Coulomb barrier and decreasing binding energies. This s-process takes place in the Red Giants
and Asymptotic Giant Branch (AGB) stars, where neutron temperature ({\it kT}) varies from 8 to 90 keV. The c.m. system Maxwellian-averaged cross sections (MACS) 
for neutron capture  are shown in Table \ref{MaxTable} and the fission values are published in the Brookhaven report.  
Addition of EAF-2010 and ROSFOND 2010 libraries \cite{10Sub,07Zab} provides 
more than 99$\%$ coverage along the {\it s-}process path. Only three values for  $^{110}$Ag and $^{132,133}$Ce 
 are not available yet because of lack of experimental data. These missing cross sections will be addressed in our future work.

$^{252}$Cf has been often used for nuclear reactor modeling. Its fission neutron spectrum data have been compiled in the EXFOR database \cite{EXFOR}. Analysis of calculated MACS and experimental data 
provides an important tool for ENDF quality assurance.

\subsubsection{{\it kT}=30 keV}

\begin{figure}[!ht]
\includegraphics[width=0.95\columnwidth]{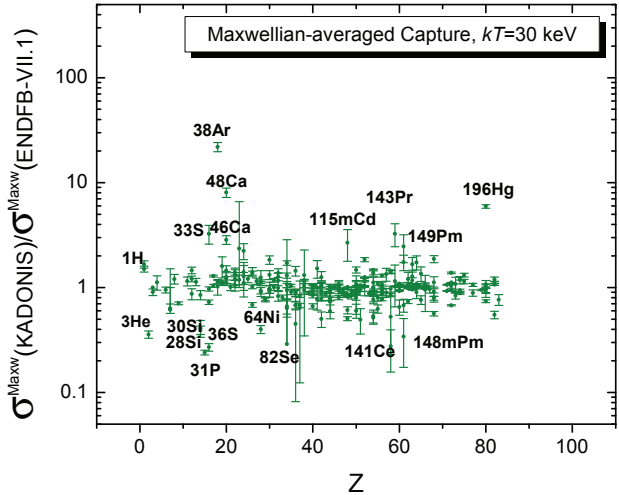}
\caption{Ratio of Maxwellian-averaged capture cross sections at {\it kT}=30 keV of Karlsruhe Astrophysical Database of Nucleosynthesis in Stars
(KADoNiS) \cite{06Dil} to ENDF/B-VII.1  calculation.}
\label{IQ.5}
\end{figure}

A detailed analysis of Fig.~\ref{IQ.5},  Table \ref{MaxTable} and the  report data demonstrates the nuclear astrophysics potential of ENDF libraries as a complementary source of evaluated cross sections and reaction rates  \cite{10Pri}. 
There are noticeable differences between KADoNiS \cite{06Dil} and ENDF/B-VII.1 libraries for light  and medium nuclei.  
The $^{1}$H deviation is due to differences between center of mass (ENDF calculated) and lab reference 
system (KADoNiS) cross section values, and therefore, this is not a discrepancy. The
$^{3}$He ENDF/B-VII.1 value is higher than KADoNiS; it comes from an R-matrix
evaluation \cite{11Chad}. For $^{28,30}$Si, $^{31}$P, $^{64}$Ni
and $^{196}$Hg the KADoNiS values are based on a single recent measurement. 
In the $^{30}$Si case, the evaluators used the higher value to tune the value, 
while the IUPAC (International Union of Pure and Applied Chemistry) constants for NAA (Neutron-Activation Analysis) support the lower value. 
 Due to lack of experimental data, theoretical values were adopted by evaluators for $^{38}$Ar, $^{82}$Se, $^{115m}$Cd, $^{141}$Ce, $^{143}$Pr and $^{148m,149}$Pm. 
Discrepancies for $^{46,48}$Ca and $^{33,36}$Si occur in part because of their small isotopic abundances and in consequence the lack of measured data \cite{EXFOR}. 
Figs.~\ref{IQ.5} and \ref{IQ.3} provide an additional proof that  $^{31}$P and $^{46}$Ca data in ENDF/B-VII.1 are not very realistic compared to available benchmarks.

\begin{figure}[htb]	
\centering
\includegraphics[width=0.95\columnwidth]{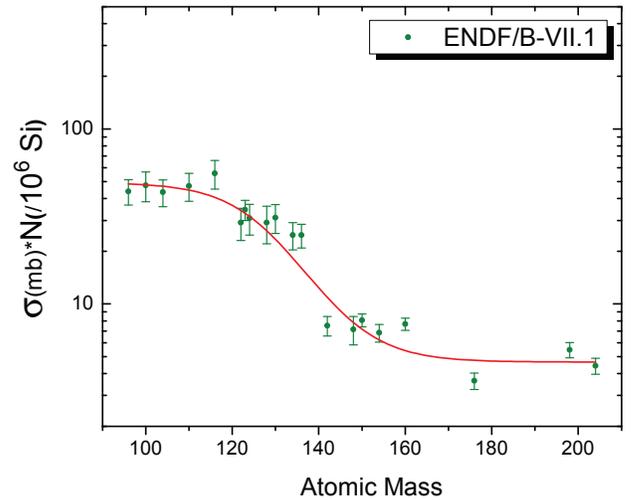} 
\caption[ENDF/B-VII.1 library product of neutron-capture cross section (at 30 keV in mb) times solar system abundances 
(relative to Si = 10$^6$) as a function of atomic mass for nuclei produced only in the {\it s}-process.]
{ENDF/B-VII.1 library product of neutron-capture cross section (at 30 keV in mb) times solar system abundances 
(relative to Si = 10$^6$) as a function of atomic mass for nuclei produced only in the {\it s}-process.}\label{sigma}
\end{figure}

In {\it s-}process nucleosynthesis, the product of neutron-capture cross section (at 30 keV in mb) times solar system abundances (relative to Si = 10$^6$) as a function of atomic 
mass should be constant for equilibrium nuclei \cite{88Rol}: 
\begin{equation} 
\label{myeq.eq} 
\sigma_{A}N_{(A)}= \sigma_{A-1}N_{(A-1)} = constant.
\end{equation}
 To verify this phenomenon, the calculated $\langle \sigma^{Maxw}_{\gamma} (30 keV) \rangle$ from the ENDF/B-VII.1 library were 
 multiplied by solar abundances taken from Anders and  Grevesse \cite{89And}, and plotted  in Fig. \ref{sigma}. 
Visual inspection of  Fig. \ref{sigma}  indicates two local equilibrium and ledge-precipice break at $A \sim$ 138 for the ENDF/B-VII.1 fit and 
relatively high  value for $^{116}$Sn. This phenomenon is due to the fact that $^{116}$Sn 
solar abundance has {\it r-}process contribution \cite{89And}.  

\begin{figure}[htb]	
\centering
\includegraphics[width=0.95\columnwidth]{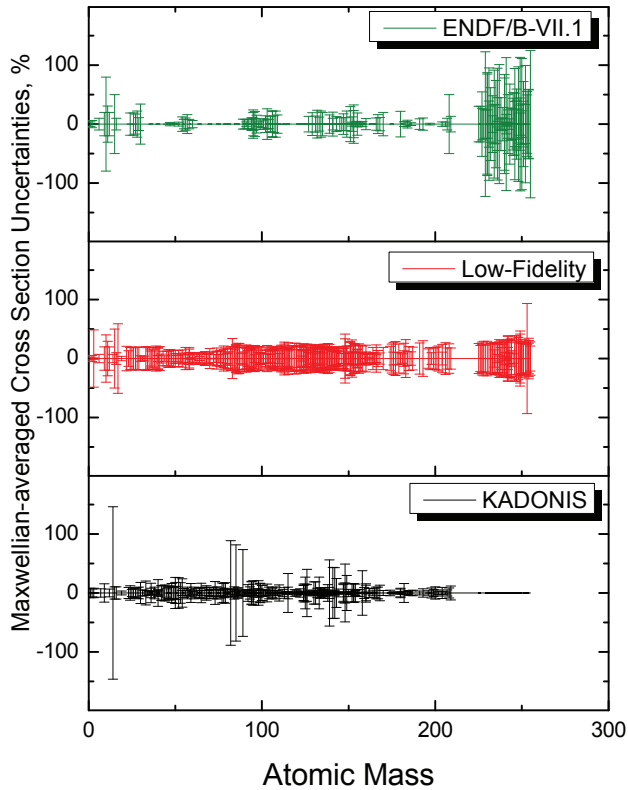} 
\caption[Maxwellian-averaged Neutron Capture Cross Sections Uncertainty for ENDF/B-VII.1 library, Low-Fidelity project and KADONIS database.]{Maxwellian-averaged Neutron Capture Cross Section Uncertainties for ENDF/B-VII.1 library, Low-Fidelity project and KADONIS database \citep{11Chad,08Lit,06Dil}.}\label{MaxUn}
\end{figure}

The predictive power of stellar nucleosynthesis calculations depends heavily on the neutron capture cross section values and their covariances. 
To understand the unique isotopic signatures from the presolar grains, $\sim$1\% cross section and decay rates uncertainties are necessary \citep{11Fka}. 
Unfortunately, present $\sigma$(n,$\gamma$) uncertainties are often much higher than 1$\%$, as shown in Fig. \ref{MaxUn}. 

An additional ENDF s-process nucleosynthesis data could be interactively calculated and  downloaded from the 
NucRates Web application {\it http://www.nndc.bnl.gov/astro}. These complimentary data sets demonstrate a strong 
correlation between nuclear astrophysics and nuclear industry data needs,  the large nuclear astrophysics potential of 
ENDF libraries, and a perspective beneficial relationship between both fields.

Analysis of fission Maxwellian-averaged cross sections, Table XI of the report, indicates potential deficiency with $^{234}$Th evaluation in 
 JEFF-3.1.2 library that lacks fission cross section data below 1.5 MeV.

\subsubsection{{\it kT}=1420 keV}

To provide the quality assurance for the high-energy part of neutron spectrum, we calculated  Maxwellian-averaged 
cross sections close to the  $^{252}$Cf spontaneous fission neutron spectrum conditions. 
$^{252}$Cf is often used in nuclear physics as a compact, portable and intense neutron source. Its alpha  
and spontaneous fission decay modes, 96.91$\%$ and 3.09$\%$, respectively, result in an overall
half-life of 2.645$\pm$0.008 years \cite{05Ni}. Its neutron emission is about 3.757 neutrons per
fission \cite{85Bol} and 1 mg of $^{252}$Cf generates $\sim$2.3$\times$10$^{9}$ neutrons/s  \cite{97Ma}.  
Californium neutron energy spectrum is similar to a fission reactor, with  an average energy of 2.13 MeV \cite{79Smi,67Me}. 
Consequently, compact $^{252}$Cf neutron sources can provide an ideal nonreactor source of neutrons for lower-flux applications.
Large masses of $^{252}$Cf ($>$l00 mg) can imitate reactor capabilities for applications such as neutron radiography. 
In the United States, Californium User Facility for Neutron Science processes and encapsulates the national 
supply of $^{252}$Cf, produced at the neighboring Oak Ridge High Flux Isotope Reactor (HFIR) \cite{252Cf}.

\begin{figure}
\begin{center}
\includegraphics[width=0.95\columnwidth]{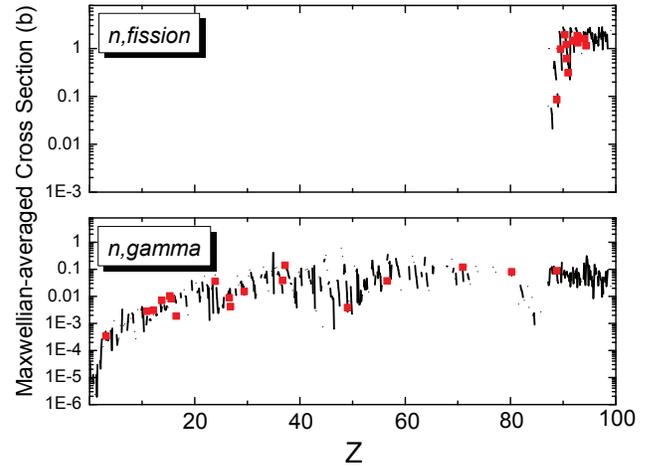}
\caption{Calculated ENDF/B-VII.1 library MACS at $kT$=1420 keV and EXFOR data for neutron-induced fission and capture reactions  are shown as solid lines and red rectangulars, respectively.}
\label{fig1420}
\end{center}
\end{figure}

Assuming the $^{252}$Cf $\bar{E}$ of 2.13 MeV  and 
employing the formula
\begin{equation}
\label{myeq.252Cf}
\bar{E}_{n} = \frac{3}{2} kT,
\end{equation}
then  $kT$=1420 keV for the approximate  Maxwellian temperature \cite{94Tru}. 
Obviously, current approximation has its limitations and the more advanced formalisms have been proposed \cite{85Gra,98Mad,10Ta}. 
However, this temperature data are consistent with the 13 (n,$\gamma$) and 17 (n,fission) EXFOR data sets \cite{EXFOR}. 
ENDF/B-VII.1 MACS combined with the EXFOR data are shown in Fig. \ref{fig1420}.
The complete set of calculated MACS at this $kT$ is given in the Tables XIII-XIV of the Brookhaven report.

\subsection{Astrophysical Reaction Rates}
\begin{figure}
\begin{center}
\includegraphics[width=0.95\columnwidth]{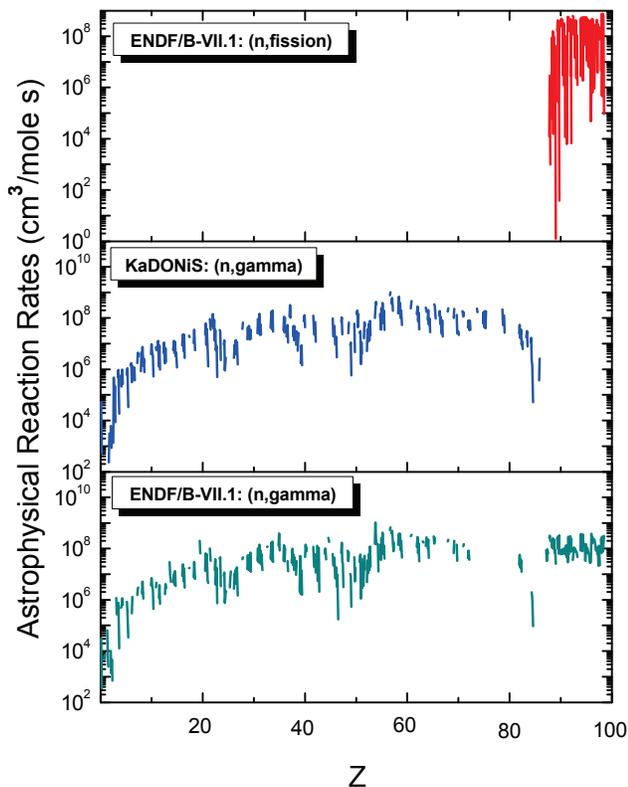}
\caption{Calculated ENDF/B-VII.1 astrophysical reaction rates at $kT$=30 keV and KaDONiS data for neutron-induced fission and capture reactions.}
\label{rr30}
\end{center}
\end{figure}

Astrophysical reaction rates and their uncertainties for the whole range of ENDF/B-VII.1 nuclei have been calculated at $kT$=30 keV and presented in the Tables XV-XVI of Brookhaven report. 
An example of ENDF/B-VII.1 reaction rates is shown in Fig. \ref{rr30}.
An additional calculation for 1 MK - 10 GK temperature range \cite{12Il} will be available in the electronic form only.  
Current calculation is in good agreement with KADONIS and other values \cite{06Mugh,06Dil,00Ra}. It practically covers the whole range of $s$-process nuclei and represents the most complete set of reaction rates produced from the 
evaluated nuclear reaction libraries.

\subsection{Low-Fidelity Project Uncertainties for ENDF/B-VII.1 Library}

To gain better understanding of ENDF/B-VII.1 library physical quantities and their uncertainties, 
the Low-Fidelity project results \cite{08Lit} were applied to calculate complete set of uncertainties. 
Data analysis of Fig. \ref{MaxUn} indicates that present uncertainties are well-suited for nuclear astrophysics  
applications and provide good coverage along $s$-process nucleosynthesis path.
These results are shown in Tables XVII-XIX of the Brookhaven report.

\subsection{Major Improvements of Present Work with Previous Evaluations}
Current work is the most complete calculation of ENDF values up-to-date \cite{05Nak,10Pri,10Sig} and 
reflects the strong USNDP commitment to the nuclear data research and development. The list of improvements include:
\begin{itemize}
\item Extended list of physical quantities: neutron thermal cross sections, Westcott factors, resonance integrals, Maxwelian-averaged cross sections for $kT$=30, 1420 keV and astrophysical reaction rates
\item Improved numerical integration for T=0, 293.6$^{\circ}$ K Doppler broadening temperatures
\item Calculation of uncertainties
\item More than 99$\%$ coverage of the $s$-process path
\item Coverage of all major libraries
\item The complete list of neutron materials that are present in major evaluated libraries
\end{itemize}

\section{CONCLUSION AND OUTLOOK}
\label{sec:conclusion}

Increasing demands for  development of new nuclear energy and astrophysics  applications provided a strong motivation for this work. 
A complete calculation of  Westcott factors, resonance integrals,  Maxwellian-averaged cross sections, astrophysical reaction rates  
and their uncertainties has been performed. Neutron thermal cross section data were extracted from the ENDF libraries. 
Present data were analyzed using benchmarks, where available. Data analysis indicates  a substantial progress in 
nuclear data libraries quality and their importance for a wide variety of applications. 


\section*{Acknowledgments}
The authors are indebted to Dr. M.W. Herman (BNL) for the constant help and support during this project and Dr. P. Oblo{\v z}insk{\' y} (Slovak Academy of Sciences) for useful suggestions. 
We acknowledge Drs. V. Zerkin (IAEA), A. Trkov (Jozef Stefan Institute, Ljubljana), R. Cyburt (JINA/MSU) and  M. Pigni (ORNL) for productive discussions
and help on ENDF Doppler broadening, Westcott factors,  nuclear astrophysics calculations and Low-Fidelity project data, respectively. We are also grateful
to  M. Blennau (BNL) for a careful reading of the manuscript. This work was funded by the Office of
Nuclear Physics, Office of Science of the U.S. Department of Energy, under Contract No. DE-AC02-98CH10886 with
Brookhaven Science Associates, LC.

\section*{Appendix}



\section*{Thermal Elastic Cross Sections}
   \newpage
   \renewcommand{\arraystretch}{0.5}
   \renewcommand\tabcolsep{1pt}
   \scriptsize
   \begin{center}

\begin{longtable*}{@{}>{\small}l|>{\small}c>{\small}c>{\small}c>{\small}c>{\small}c>{\small}c>{\small}c>{\small}c>{\small}c@{}}
   \caption{Thermal elastic cross sections  for T = 0, 293.6$^{\circ}$ processing temperatures from ENDF/B-VII.1, JEFF-3.1.2, JENDL-4.0, ROSFOND-2010, CENDL-3.1,  EAF-2010, Atlas of Neutron Resonances and Neutron Cross Section Standards \cite{11Chad,11Kon,11Shi,07Zab,11Zhi,10Sub,06Mugh,09Ca}.} \label{Therm2Table}
   \\
   \hline \hline \\
   \textbf{Material} & \textbf{T} & \textbf{ENDF/B} & \textbf{JEFF} & \textbf{JENDL} & \textbf{ROSFOND} & \textbf{CENDL}  & \textbf{EAF} & \textbf{Atlas} & \textbf{Standards}\\
   	& (K) & (barns) & (barns) & (barns) & (barns) & (barns) & (barns) & (barns) & (barns)	\\
   \hline \\
   \endfirsthead
   \caption[]{Thermal elastic cross sections  (continued).}
   \\
   \hline \hline \\
   \textbf{Material} & \textbf{T} & \textbf{ENDF/B} & \textbf{JEFF} & \textbf{JENDL} &  \textbf{ROSFOND} & \textbf{CENDL}  & \textbf{EAF} & \textbf{Atlas} & \textbf{Standards}\\
   	& (K) & (barns) & (barns) & (barns) & (barns) & (barns) & (barns) & (barns) & (barns)	\\
   \hline \\
\endhead
 1-H -  1 & 0 & 2.040E+1 & 2.050E+1 & 2.040E+1 & 2.040E+1 & 2.050E+1 &  & 2.050E+1$\pm$1.400E-2 & 20.43633$\pm$0.04087266 \\
 & 293.6 & 3.010E+1 & 3.010E+1 & 3.000E+1 & 3.000E+1 & 3.010E+1 &  &  &  \\
 1-H -  2 & 0 & 3.400E+0 & 3.400E+0 & 3.390E+0 & 3.390E+0 & 3.400E+0 &  & 3.390E+0$\pm$1.200E-2 &  \\
 & 293.6 & 4.240E+0 & 4.240E+0 & 4.230E+0 & 4.230E+0 & 4.240E+0 &  &  &  \\
 1-H -  3 & 0 & 1.670E+0 & 1.700E+0 &  & 1.700E+0 & 1.700E+0 &  & 1.700E+0$\pm$3.000E-2 &  \\
 & 293.6 & 1.940E+0 & 1.980E+0 &  & 1.980E+0 & 1.980E+0 &  &  &  \\
 2-He-  3 & 0 & 3.310E+0 & 3.310E+0 & 3.140E+0 & 3.160E+0 & 3.140E+0 &  & 3.100E+0$\pm$1.300E-1 &  \\
 & 293.6 & 3.860E+0 & 3.860E+0 & 3.660E+0 & 3.690E+0 & 3.660E+0 &  &  &  \\
\hline \hline
 \end{longtable*}

 \end{center}
\normalsize

\section*{Thermal Fission Cross Sections}
  \newpage
  \renewcommand{\arraystretch}{0.5}
  \renewcommand\tabcolsep{1pt}
  \scriptsize
  \begin{center}

  \begin{longtable*}{@{}>{\small}l|>{\small}c>{\small}c>{\small}c>{\small}c>{\small}c>{\small}c>{\small}c>{\small}c@{}}
  \caption{Thermal fission cross sections from ENDF/B-VII.1, JEFF-3.1.2, JENDL-4.0, ROSFOND-2010, CENDL-3.1,  EAF-2010, Atlas of Neutron Resonances and Neutron Cross Section Standards \cite{11Chad,11Kon,11Shi,07Zab,11Zhi,10Sub,06Mugh,09Ca}.} \label{Therm18Table}
  \\
  \hline \hline \\
  \textbf{Material} & \textbf{ENDF/B} & \textbf{JEFF} & \textbf{JENDL} & \textbf{ROSFOND} & \textbf{CENDL}  & \textbf{EAF} & \textbf{Atlas} & \textbf{Standards}\\
  	& (barns) & (barns) & (barns) & (barns) & (barns) & (barns) & (barns) & (barns)	\\
  \hline \\
  \endfirsthead
  \caption[]{Thermal fission cross sections (continued).}
  \\
  \hline \hline \\
  \textbf{Material} & \textbf{ENDF/B} & \textbf{JEFF} & \textbf{JENDL} &  \textbf{ROSFOND} & \textbf{CENDL}  & \textbf{EAF} & \textbf{Atlas} & \textbf{Standards}\\
  	& (barns) & (barns) & (barns) & (barns) & (barns) & (barns) & (barns) & (barns)	\\
  \hline \\
\endhead

 86-Rn-222 &  &  &  & 6.202E-6 &  & 1.000E-1 &  & \\
 88-Ra-223 & 7.003E-1 & 7.003E-1 & 7.003E-1 & 7.003E-1 &  & 7.000E-1 &  & \\
 88-Ra-224 &  &  &  &  &  & 1.800E-3 &  & \\
 88-Ra-225 &  &  &  &  &  & 1.000E+1 &  & \\
 88-Ra-226 & 7.002E-6 & 7.002E-6 & 7.002E-6 & 7.002E-6 &  & 7.000E-6 &  & \\
 88-Ra-228 &  &  &  & 2.001E+0 &  & 2.000E+0 & $<$2 & \\
 89-Ac-225 & 5.003E-4 &  & 5.003E-4 & 1.801E-3 &  & 1.800E-3 &  & \\
 89-Ac-226 & 2.001E+0 &  & 2.001E+0 & 6.703E+2 &  & 8.000E+0 &  & \\
\hline \hline
 \end{longtable*}

 \end{center}
\normalsize

\section*{Thermal Capture Cross Sections}
  \newpage
  \renewcommand{\arraystretch}{0.5}
  \renewcommand\tabcolsep{1pt}
  \scriptsize
  \begin{center}

  \begin{longtable*}{@{}>{\small}l|>{\small}c>{\small}c>{\small}c>{\small}c>{\small}c>{\small}c>{\small}c>{\small}c@{}}
  \caption{Thermal capture cross sections from ENDF/B-VII.1, JEFF-3.1.2, JENDL-4.0, ROSFOND-2010, CENDL-3.1,  EAF-2010, Atlas of Neutron Resonances and Neutron Cross Section Standards \cite{11Chad,11Kon,11Shi,07Zab,11Zhi,10Sub,06Mugh,09Ca}.} \label{Therm102Table}
  \\
  \hline \hline \\
  \textbf{Material} & \textbf{ENDF/B} & \textbf{JEFF} & \textbf{JENDL} & \textbf{ROSFOND} & \textbf{CENDL}  & \textbf{EAF} & \textbf{Atlas} & \textbf{Standards}\\
  	& (barns) & (barns) & (barns) & (barns) & (barns) & (barns) & (barns) & (barns)	\\
  \hline \\
  \endfirsthead
  \caption[]{Thermal capture cross sections (continued).}
  \\
  \hline \hline \\
  \textbf{Material} & \textbf{ENDF/B} & \textbf{JEFF} & \textbf{JENDL} &  \textbf{ROSFOND} & \textbf{CENDL}  & \textbf{EAF} & \textbf{Atlas} & \textbf{Standards}\\
  	& (barns) & (barns) & (barns) & (barns) & (barns) & (barns) & (barns) & (barns)	\\
  \hline \\
\endhead
 1-H -  1 & 3.320E-1 & 3.320E-1 & 3.320E-1 & 3.320E-1 & 3.320E-1 & 3.320E-1 & 3.326E-1$\pm$7.000E-4 &  \\
 1-H -  2 & 5.057E-4 & 5.057E-4 & 5.500E-4 & 5.190E-4 & 5.057E-4 & 5.057E-4 & 5.080E-4$\pm$1.500E-5 &  \\
 1-H -  3 &  &  &  &  &  &  & $<$0.000006 &  \\
 2-He-  3 & 5.499E-5 & 3.100E-5 & 5.402E-5 & 2.201E-5 & 5.402E-5 & 5.402E-5 & 5.500E-5$\pm$3.000E-6 &  \\
 2-He-  4 &  &  &  &  &  &  \\
 3-Li-  6 & 3.850E-2 & 3.850E-2 & 3.849E-2 & 3.850E-2 &  & 3.853E-2 & 3.850E-2$\pm$3.000E-3 &  \\
 3-Li-  7 & 4.540E-2 & 4.540E-2 & 4.540E-2 & 4.540E-2 & 4.540E-2 & 4.545E-2 & 4.540E-2$\pm$2.700E-3 &  \\
 4-Be-  7 &  &  &  &  &  & 2.509E-2 \\
\hline \hline
\end{longtable*}

\end{center}
\normalsize

\section*{Westcott Elastic Factors}
  \newpage
  \renewcommand{\arraystretch}{0.5}
  \renewcommand\tabcolsep{1pt}
  \scriptsize
  \begin{center}

  \begin{longtable*}{@{}>{\small}l|>{\small}c>{\small}c>{\small}c>{\small}c>{\small}c>{\small}c@{}}
  \caption{Westcott neutron elastic factors at T=293.6$^{\circ}$ K  from ENDF/B-VII.1, JEFF-3.1.2, JENDL-4.0, ROSFOND-2010, CENDL-3.1  libaries and EAF-2010 file  \cite{11Chad,11Kon,11Shi,07Zab,11Zhi,10Sub}.} \label{West2Table}
  \\
  \hline \hline \\
  \textbf{Material} & \textbf{ENDF/B} & \textbf{JEFF} & \textbf{JENDL} & \textbf{ROSFOND} & \textbf{CENDL}  & \textbf{EAF}  \\
  	&  &  &  &  &  & 	\\
  \hline \\
  \endfirsthead
  \caption[]{Westcott neutron elastic factors  (continued).}
  \\
  \hline \hline \\
  \textbf{Material} & \textbf{ENDF/B} & \textbf{JEFF} & \textbf{JENDL} &  \textbf{ROSFOND} & \textbf{CENDL}  & \textbf{EAF}  \\
  	&  &  &  &  &  &   \\
  \hline \\
\endhead 
 1-H -  1 & 1.085E+0 & 1.085E+0 & 1.084E+0 & 1.084E+0 & 1.085E+0 &  \\
 1-H -  2 & 1.109E+0 & 1.109E+0 & 1.109E+0 & 1.109E+0 & 1.109E+0 &  \\
 1-H -  3 & 1.118E+0 & 1.118E+0 &  & 1.117E+0 & 1.117E+0 &  \\
 2-He-  3 & 1.118E+0 & 1.118E+0 & 1.118E+0 & 1.118E+0 & 1.118E+0 &  \\
 2-He-  4 & 1.122E+0 & 1.122E+0 & 1.122E+0 & 1.122E+0 & 1.122E+0 &  \\
 3-Li-  6 & 1.126E+0 & 1.126E+0 & 1.126E+0 & 1.126E+0 & 1.126E+0 &  \\
 3-Li-  7 & 1.127E+0 & 1.127E+0 & 1.127E+0 & 1.128E+0 & 1.120E+0 &  \\
 4-Be-  7 & 1.126E+0 &  &  &  &  &  \\
\hline \hline
\end{longtable*}

\end{center}
\normalsize

\section*{Westcott Fission Factors}
  \newpage
  \renewcommand{\arraystretch}{0.5}
  \renewcommand\tabcolsep{1pt}
  \scriptsize
  \begin{center}

  \begin{longtable*}{@{}>{\small}l|>{\small}c>{\small}c>{\small}c>{\small}c>{\small}c>{\small}c>{\small}c@{}}
  \caption{Westcott neutron fission factors at T=293.6$^{\circ}$ K  from ENDF/B-VII.1, JEFF-3.1.2, JENDL-4.0, ROSFOND-2010, CENDL-3.1 libaries, EAF-2010 file and Atlas of Neutron Resonances  \cite{11Chad,11Kon,11Shi,07Zab,11Zhi,10Sub,06Mugh}.} \label{West18Table}
  \\
  \hline \hline \\
  \textbf{Material} & \textbf{ENDF/B} & \textbf{JEFF} & \textbf{JENDL} & \textbf{ROSFOND} & \textbf{CENDL}  & \textbf{EAF}  & \textbf{Atlas}  \\
  	&  &  &  &  &  &  & 	\\
  \hline \\
  \endfirsthead
  \caption[]{Westcott neutron fission factors  (continued).}
  \\
  \hline \hline \\
  \textbf{Material} & \textbf{ENDF/B} & \textbf{JEFF} & \textbf{JENDL} &  \textbf{ROSFOND} & \textbf{CENDL}  & \textbf{EAF} & \textbf{Atlas} \\
  	&  &  &  &  &  &  &  \\
  \hline \\
\endhead 
 86-Rn-222 &  &  &  & 1.001E+0 &  & 9.995E-1 &  \\
 88-Ra-223 & 1.001E+0 & 1.001E+0 & 1.001E+0 & 1.001E+0 &  & 9.996E-1 &  \\
 88-Ra-224 &  &  &  &  &  & 9.995E-1 &  \\
 88-Ra-225 &  &  &  &  &  & 9.993E-1 &  \\
 88-Ra-226 & 1.000E+0 & 1.000E+0 & 1.000E+0 & 1.000E+0 &  & 9.991E-1 &  \\
 88-Ra-228 &  &  &  & 1.001E+0 &  & 9.995E-1 &  \\
 89-Ac-225 & 1.002E+0 &  & 1.002E+0 & 1.001E+0 &  & 9.995E-1 &  \\
 89-Ac-226 & 1.001E+0 &  & 1.001E+0 & 1.000E+0 &  & 9.993E-1 &  \\
\hline \hline
 \end{longtable*}

 \end{center}
\normalsize

\section*{Westcott Capture Factors}
  \newpage
  \renewcommand{\arraystretch}{0.5}
  \renewcommand\tabcolsep{1pt}
  \scriptsize
  \begin{center}

  \begin{longtable*}{@{}>{\small}l|>{\small}c>{\small}c>{\small}c>{\small}c>{\small}c>{\small}c>{\small}c@{}}
  \caption{Westcott neutron capture factors at T=293.6$^{\circ}$ K  from ENDF/B-VII.1, JEFF-3.1.2, JENDL-4.0, ROSFOND-2010, CENDL-3.1,  EAF-2010 and  Atlas of Neutron Resonances \cite{11Chad,11Kon,11Shi,07Zab,11Zhi,10Sub,06Mugh}. ({\it C}-calculated value).} \label{West102Table}
  \\
  \hline \hline \\
  \textbf{Material} & \textbf{ENDF/B} & \textbf{JEFF} & \textbf{JENDL} & \textbf{ROSFOND} & \textbf{CENDL}  & \textbf{EAF} & \textbf{Atlas} \\
  	&  &  &  &  &  &  &  	\\
  \hline \\
  \endfirsthead
  \caption[]{Westcott neutron capture factors  (continued).}
  \\
  \hline \hline \\
  \textbf{Material} & \textbf{ENDF/B} & \textbf{JEFF} & \textbf{JENDL} &  \textbf{ROSFOND} & \textbf{CENDL}  & \textbf{EAF} & \textbf{Atlas} \\
  	&  &  &  &  &  &  &  \\
  \hline \\
\endhead

 1-H -  1 & 1.001E+0 & 1.001E+0 & 1.001E+0 & 1.001E+0 & 1.001E+0 & 9.995E-1 &  \\
 1-H -  2 & 1.000E+0 & 1.000E+0 & 1.001E+0 & 1.001E+0 & 1.000E+0 & 9.993E-1 &  \\
 1-H -  3 &  &  &  &  &  &  &  \\
 2-He-  3 & 1.001E+0 & 1.001E+0 & 1.001E+0 & 9.851E-1 & 1.001E+0 & 9.996E-1 &  \\
 2-He-  4 &  &  &  &  &  &  &  \\
 3-Li-  6 & 1.001E+0 & 1.001E+0 & 1.000E+0 & 1.001E+0 &  & 9.987E-1 &  \\
 3-Li-  7 & 1.001E+0 & 1.001E+0 & 1.001E+0 & 1.001E+0 & 1.002E+0 & 9.988E-1 &  \\
 4-Be-  7 &  &  &  &  &  & 9.995E-1 &  \\
\hline \hline
 \end{longtable*}

 \end{center}
\normalsize



\section*{Elastic Resonance Integrals}
   \newpage
   \renewcommand{\arraystretch}{0.5}
   \renewcommand\tabcolsep{1pt}
   \scriptsize
   \begin{center}

   \begin{longtable*}{@{}>{\small}l|>{\small}c>{\small}c>{\small}c>{\small}c>{\small}c>{\small}c@{}}
   \caption{Elastic Resonance Integrals from ENDF/B-VII.1, JEFF-3.1.2, JENDL-4.0, ROSFOND-2010. CENDL-3.1 libraris and EAF-2010 file \cite{11Chad,11Kon,11Shi,07Zab,11Zhi,10Sub}.} \label{RI2Table}
   \\
   \hline \hline \\
   \textbf{Material} & \textbf{ENDF/B} & \textbf{JEFF} & \textbf{JENDL} & \textbf{ROSFOND} & \textbf{CENDL}  & \textbf{EAF}  \\
   	& (barns) & (barns) & (barns) & (barns) & (barns) & (barns) 	\\
   \hline \\
   \endfirsthead
   \caption[]{Elastic Resonance Integrals (continued).}
   \\
   \hline \hline \\
   \textbf{Material} & \textbf{ENDF/B} & \textbf{JEFF} & \textbf{JENDL} &  \textbf{ROSFOND} & \textbf{CENDL}  & \textbf{EAF}  \\
   	& (barns) & (barns) & (barns) & (barns) & (barns) & (barns) 	\\
   \hline \\
\endhead

 1-H -  1 & 2.645E+2 & 2.647E+2 & 2.619E+2 & 2.619E+2 & 2.648E+2 & \\
 1-H -  2 & 5.342E+1 & 5.342E+1 & 5.335E+1 & 5.336E+1 & 5.346E+1 & \\
 1-H -  3 & 2.933E+1 & 2.951E+1 &  & 2.957E+1 & 2.951E+1 & \\
 2-He-  3 & 4.871E+1 & 4.871E+1 & 4.662E+1 & 4.758E+1 & 4.657E+1 & \\
 2-He-  4 & 2.216E+1 & 2.199E+1 & 2.199E+1 & 2.199E+1 & 2.201E+1 & \\
 3-Li-  6 & 1.782E+1 & 1.727E+1 & 1.790E+1 & 1.756E+1 & 1.756E+1 & \\
 3-Li-  7 & 2.135E+1 & 2.135E+1 & 2.092E+1 & 2.135E+1 & 2.202E+1 & \\
 4-Be-  7 & 2.054E+2 &  &  &  &  & \\
\hline \hline
 \end{longtable*}

 \end{center}
\normalsize

\section*{Fission Resonance Integrals}
\newpage
\renewcommand{\arraystretch}{0.5}
\renewcommand\tabcolsep{1pt}
\scriptsize
\begin{center}

\begin{longtable*}{@{}>{\small}l|>{\small}c>{\small}c>{\small}c>{\small}c>{\small}c>{\small}c>{\small}c@{}}
\caption{Fission Resonance Integrals from ENDF/B-VII.1, JEFF-3.1.2, JENDL-4.0, ROSFOND-2010, CENDL-3.1,  EAF-2010 and Atlas of Neutron Resonances \cite{11Chad,11Kon,11Shi,07Zab,11Zhi,10Sub,06Mugh}, ({\it C}-calculated values).} \label{RI18Table}
 \\
\hline \hline \\
\textbf{Material} & \textbf{ENDF/B} & \textbf{JEFF} & \textbf{JENDL} & \textbf{ROSFOND} & \textbf{CENDL}  & \textbf{EAF} & \textbf{Atlas} \\
   	& (barns) & (barns) & (barns) & (barns) & (barns) & (barns) & (barns)	\\
\hline \\
\endfirsthead
\caption[]{Fission Resonance Integrals  (continued).}
\\
\hline \hline \\
\textbf{Material} & \textbf{ENDF/B} & \textbf{JEFF} & \textbf{JENDL} &  \textbf{ROSFOND} & \textbf{CENDL}  & \textbf{EAF} & \textbf{Atlas} \\
   	& (barns) & (barns) & (barns) & (barns) & (barns) & (barns) & (barns)	\\
\hline \\
\endhead

 86-Rn-222 &  &  &  & 8.756E-1 &  & 6.179E-1 & \\
 88-Ra-223 & 1.062E+0 & 1.062E+0 & 1.062E+0 & 1.062E+0 &  & 1.065E+0 & \\
 88-Ra-224 &  &  &  &  &  & 8.765E-1 & \\
 88-Ra-225 &  &  &  &  &  & 5.624E+0 & \\
 88-Ra-226 & 1.187E-2 & 1.187E-2 & 1.187E-2 & 1.187E-2 &  & 1.223E-2 & \\
 88-Ra-228 &  &  &  & 1.125E+0 &  & 1.125E+0 & \\
 89-Ac-225 & 2.159E-1 &  & 2.159E-1 & 8.764E-1 &  & 8.765E-1 & \\
 89-Ac-226 & 8.705E+0 &  & 8.705E+0 & 3.027E+2 &  & 5.337E+0 & \\
\hline \hline
  \end{longtable*}

  \end{center}
\normalsize

\section*{Capture Resonance Integrals}
\newpage
\renewcommand{\arraystretch}{0.5}
\renewcommand\tabcolsep{1pt}
\scriptsize
\begin{center}

\begin{longtable*}{@{}>{\small}l|>{\small}c>{\small}c>{\small}c>{\small}c>{\small}c>{\small}c>{\small}c@{}}
\caption{Capture Resonance Integrals from ENDF/B-VII.1, JEFF-3.1.2, JENDL-4.0, ROSFOND-2010, CENDL-3.1,  EAF-2010 and Atlas of Neutron Resonances \cite{11Chad,11Kon,11Shi,07Zab,11Zhi,10Sub,06Mugh}, ({\it C}-calculated values).} \label{RI102Table}
 \\
\hline \hline \\
\textbf{Material} & \textbf{ENDF/B} & \textbf{JEFF} & \textbf{JENDL} & \textbf{ROSFOND} & \textbf{CENDL}  & \textbf{EAF} & \textbf{Atlas} \\
   	& (barns) & (barns) & (barns) & (barns) & (barns) & (barns) & (barns)	\\
\hline \\
\endfirsthead
\caption[]{Capture Resonance Integrals  (continued).}
\\
\hline \hline \\
\textbf{Material} & \textbf{ENDF/B} & \textbf{JEFF} & \textbf{JENDL} &  \textbf{ROSFOND} & \textbf{CENDL}  & \textbf{EAF} & \textbf{Atlas} \\
   	& (barns) & (barns) & (barns) & (barns) & (barns) & (barns) & (barns)	\\
\hline \\
\endhead 
 1-H -  1 & 1.492E-1 & 1.492E-1 & 1.492E-1 & 1.492E-1 & 1.515E-1 & 1.491E-1 & \\
 1-H -  2 & 2.640E-4 & 2.640E-4 & 2.865E-4 & 2.774E-4 & 2.635E-4 & 2.642E-4 & \\
 1-H -  3 &  &  &  &  &  &  & \\
 2-He-  3 & 3.637E-4 & 1.396E-5 & 2.468E-4 & 3.577E-4 & 2.468E-4 & 2.457E-4 & \\
 2-He-  4 &  &  &  &  &  &  & \\
 3-Li-  6 & 1.736E-2 & 1.736E-2 & 1.749E-2 & 1.736E-2 &  & 1.748E-2 & 1.730E-2$\pm$1.400E-3$^{C}$ \\
 3-Li-  7 & 2.045E-2 & 2.045E-2 & 2.044E-2 & 2.045E-2 & 2.660E-2 & 2.045E-2 & 2.200E-2$\pm$2.000E-3$^{C}$ \\
 4-Be-  7 &  &  &  &  &  & 1.974E-2 & \\
\hline \hline
  \end{longtable*}

  \end{center}
\normalsize

\section*{Maxwellian-averaged 30-keV Capture Cross Sections}
  \newpage
  \renewcommand{\arraystretch}{0.5}
  \renewcommand\tabcolsep{1pt}
  \scriptsize
  \begin{center}

  \begin{longtable*}{@{}>{\small}l|>{\small}c>{\small}c>{\small}c>{\small}c>{\small}c>{\small}c>{\small}c@{}}
  \caption{Maxwellian-averaged cross sections from ENDF/B-VII.1, JEFF-3.1.2, JENDL-4.0, ROSFOND-2010, CENDL-3.1,  EAF-2010 and  KADoNiS \cite{11Chad,11Kon,11Shi,07Zab,11Zhi,10Sub,06Dil}  at {\it kT}=30 keV ({\it C}-calculated from BNL-325 data \cite{81Mu}, {\it T}-theoretical data in KADoNiS). {\it G}-ground state value, $^*$~Updated values for 40-Zr- 95, 69-Tm-169 and 69-Tm-170.} \label{MaxTable}
  \\
  \hline \hline \\
  \textbf{Material} & \textbf{ENDF/B-VII.1} & \textbf{JEFF-3.1.2} & \textbf{JENDL-4.0} & \textbf{ROSFOND} & \textbf{CENDL-3.1}  & \textbf{EAF-2010} & \textbf{KADoNiS}\\
  	& (barns) & (barns) & (barns) & (barns) & (barns) & (barns) & (barns)	\\
  \hline \\
  \endfirsthead
  \caption[]{Maxwellian-averaged cross sections (continued).}
  \\
  \hline \hline \\
  \textbf{Material} & \textbf{ENDF/B-VII.1} & \textbf{JEFF-3.1.2} & \textbf{JENDL-4.0} &  \textbf{ROSFOND} & \textbf{CENDL-3.1}  & \textbf{EAF-2010} & \textbf{KADoNiS}\\
  	& (barns) & (barns) & (barns) & (barns) & (barns) & (barns) & (barns)	\\
  \hline \\
\endhead

  1-H -  1 &  1.525E-4$\pm$5.855E-6  & 1.524E-4 & 1.525E-4 & 1.525E-4 & 1.575E-4 & 1.523E-4 &  2.540E-4$\pm$2.000E-5  \\
  1-H -  2 &  1.998E-6$\pm$1.265E-7  & 1.998E-6 & 2.361E-6 & 1.999E-6 & 1.996E-6 & 1.999E-6 &  3.000E-6$\pm$2.000E-7  \\
  1-H -  3 &    &    &  &    &    &    &    \\
  2-He-  3 & 2.126E-5 & 2.442E-8 & 1.640E-5 & 2.316E-5 & 1.640E-5 & 1.640E-5 &  7.600E-6$\pm$6.000E-7  \\
  2-He-  4 &    &    &    &    &    &    &    \\
  3-Li-  6 &  3.276E-5$\pm$3.181E-6  & 3.276E-5 & 3.333E-5 & 3.275E-5 &    & 3.333E-5 &    \\
  3-Li-  7 &  4.645E-5$\pm$2.103E-5  & 4.645E-5 & 3.902E-5 & 4.645E-5 & 5.344E-5 & 4.645E-5 &  4.200E-5$\pm$3.000E-6  \\
  4-Be-  7 &    &  &  &    &  & 2.958E-5 &    \\
\hline \hline
\end{longtable*}

\end{center}
\normalsize

\section*{Maxwellian-averaged 30-keV Fission Cross Sections}
  \newpage
  \renewcommand{\arraystretch}{0.5}
  \renewcommand\tabcolsep{1pt}
  \scriptsize
  \begin{center}

  \begin{longtable*}{@{}>{\small}l|>{\small}c>{\small}c>{\small}c>{\small}c>{\small}c>{\small}c@{}}
  \caption{Maxwellian-averaged neutron fission cross sections from ENDF/B-VII.1, JEFF-3.1.2, JENDL-4.0, ROSFOND-2010, CENDL-3.1  and  EAF-2010  \cite{11Chad,11Kon,11Shi,07Zab,11Zhi,10Sub}  at {\it kT}=30 keV.} \label{MaxTable18}
  \\
  \hline \hline \\
  \textbf{Material} & \textbf{ENDF/B-VII.1} & \textbf{JEFF-3.1.2} & \textbf{JENDL-4.0} & \textbf{ROSFOND-2010} & \textbf{CENDL-3.1}  & \textbf{EAF-2010} \\
  	& (barns) & (barns) & (barns) & (barns) & (barns) & (barns) 	\\
  \hline \\
  \endfirsthead
  \caption[]{Maxwellian-averaged neutron fission cross sections (continued).}
  \\
  \hline \hline \\
  \textbf{Material} & \textbf{ENDF/B-VII.1} & \textbf{JEFF-3.1.2} & \textbf{JENDL-4.0} &  \textbf{ROSFOND-2010} & \textbf{CENDL-3.1}  & \textbf{EAF-2010} \\
  	& (barns) & (barns) & (barns) & (barns) & (barns) & (barns) 	\\
  \hline \\
\endhead
 86-Rn-222 &  &  &  & 3.500E-8 &  & 3.197E-4 \\
 88-Ra-223 & 6.284E-2 & 6.284E-2 & 6.284E-2 & 6.284E-2 &  & 6.284E-2 \\
 88-Ra-224 &  &  &  &  &  & 1.022E-5 \\
 88-Ra-225 &  &  &  &  &  & 5.677E-2 \\
 88-Ra-226 & 3.949E-7 & 3.949E-7 & 3.949E-7 & 3.949E-7 &  & 3.950E-7 \\
 88-Ra-228 &  &  &  & 1.135E-2 &  & 1.135E-2 \\
 89-Ac-225 & 8.471E-5$\pm$1.979E-4 &  & 8.471E-5 & 1.022E-5 &  & 1.022E-5 \\
 89-Ac-226 & 1.977E-2$\pm$4.415E-2 &  & 1.977E-2 & 6.220E-1 &  & 1.160E-1 \\
\hline \hline
\end{longtable*}

\end{center}
\normalsize

\section*{Maxwellian-averaged 1420-keV Fission Cross Sections}
  \newpage
  \renewcommand{\arraystretch}{0.5}
  \renewcommand\tabcolsep{1pt}
  \scriptsize
  \begin{center}

  \begin{longtable*}{@{}>{\small}l|>{\small}c>{\small}c>{\small}c>{\small}c>{\small}c>{\small}c>{\small}c@{}}
  \caption{Maxwellian-averaged fission cross sections from ENDF/B-VII.1, JEFF-3.1.2, JENDL-4.0, ROSFOND-2010, CENDL-3.1,  EAF-2010 and  EXFOR \cite{11Chad,11Kon,11Shi,07Zab,11Zhi,10Sub,EXFOR}  at {\it kT}=1420 keV.} \label{MaxTable142018}
  \\
  \hline \hline \\
  \textbf{Material} & \textbf{ENDF/B-VII.1} & \textbf{JEFF-3.1.2} & \textbf{JENDL-4.0} & \textbf{ROSFOND-2010} & \textbf{CENDL-3.1}  & \textbf{EAF-2010} & \textbf{EXFOR}\\
  	& (barns) & (barns) & (barns) & (barns) & (barns) & (barns) & (barns)	\\
  \hline \\
  \endfirsthead
  \caption[]{Maxwellian-averaged fission cross sections (continued).}
  \\
  \hline \hline \\
  \textbf{Material} & \textbf{ENDF/B-VII.1} & \textbf{JEFF-3.1.2} & \textbf{JENDL-4.0} &  \textbf{ROSFOND-2010} & \textbf{CENDL-3.1}  & \textbf{EAF-2010} & \textbf{EXFOR}\\
  	& (barns) & (barns) & (barns) & (barns) & (barns) & (barns) & (barns)	\\
  \hline \\
\endhead
 86-Rn-222 &  &  &  & 2.080E-1 &  & 5.893E-2 &  \\
 88-Ra-223 & 6.285E-2 & 6.285E-2 & 6.285E-2 & 6.285E-2 &  & 6.288E-2 &  \\
 88-Ra-224 &  &  &  &  &  & 2.080E-1 &  \\
 88-Ra-225 &  &  &  &  &  & 2.169E-1 &  \\
 88-Ra-226 & 7.884E-4 & 7.884E-4 & 7.884E-4 & 7.884E-4 &  & 7.911E-4 &  \\
 88-Ra-228 &  &  &  & 4.337E-2 &  & 4.337E-2 &  \\
 89-Ac-225 & 5.559E-2$\pm$5.760E-2 &  & 5.559E-2 & 2.080E-1 &  & 2.080E-1 &  \\
 89-Ac-226 & 4.237E-2$\pm$4.351E-2 &  & 4.237E-2 & 2.677E-1 &  & 2.668E-1 &  \\
\hline \hline
\end{longtable*}

\end{center}
\normalsize

\section*{Maxwellian-averaged 1420-keV Capture Cross Sections}
  \newpage
  \renewcommand{\arraystretch}{0.5}
  \renewcommand\tabcolsep{1pt}
  \scriptsize
  \begin{center}

  \begin{longtable*}{@{}>{\small}l|>{\small}c>{\small}c>{\small}c>{\small}c>{\small}c>{\small}c>{\small}c@{}}
  \caption{Maxwellian-averaged capture cross sections from ENDF/B-VII.1, JEFF-3.1.2, JENDL-4.0, ROSFOND-2010, CENDL-3.1,  EAF-2010 and  EXFOR \cite{11Chad,11Kon,11Shi,07Zab,11Zhi,10Sub,EXFOR}  at {\it kT}=1420 keV.} \label{MaxTable1420}
  \\
  \hline \hline \\
  \textbf{Material} & \textbf{ENDF/B-VII.1} & \textbf{JEFF-3.1.2} & \textbf{JENDL-4.0} & \textbf{ROSFOND-2010} & \textbf{CENDL-3.1}  & \textbf{EAF-2010} & \textbf{EXFOR}\\
  	& (barns) & (barns) & (barns) & (barns) & (barns) & (barns) & (barns)	\\
  \hline \\
  \endfirsthead
  \caption[]{Maxwellian-averaged capture cross sections (continued).}
  \\
  \hline \hline \\
  \textbf{Material} & \textbf{ENDF/B-VII.1} & \textbf{JEFF-3.1.2} & \textbf{JENDL-4.0} &  \textbf{ROSFOND-2010} & \textbf{CENDL-3.1}  & \textbf{EAF-2010} & \textbf{EXFOR}\\
  	& (barns) & (barns) & (barns) & (barns) & (barns) & (barns) & (barns)	\\
  \hline \\
\endhead
 1-H -  1 & 3.892E-5$\pm$1.460E-5 & 3.923E-5 & 3.890E-5 & 3.890E-5 & 3.812E-5 & 3.929E-5 &  \\
 1-H -  2 & 1.025E-5$\pm$7.906E-6 & 1.025E-5 & 9.994E-6 & 1.024E-5 & 9.977E-6 & 1.026E-5 &  \\
 1-H -  3 &  &  &  &  &  &  &  \\
 2-He-  3 & 8.244E-5 & 7.656E-9 & 4.781E-5 & 8.220E-5 & 4.781E-5 & 4.783E-5 &  \\
 2-He-  4 &  &  &  &  &  &  &  \\
 3-Li-  6 & 1.253E-5$\pm$9.910E-6 & 1.253E-5 & 4.273E-5 & 1.253E-5 &  & 4.283E-5 &  \\
 3-Li-  7 & 5.928E-6 & 5.928E-6 & 3.902E-5 & 5.930E-6 & 6.404E-6 & 5.927E-6 &  \\
 4-Be-  7 &  &  &  &  &  & 2.650E-3 &  \\
\hline \hline
 \end{longtable*}

 \end{center}
\normalsize

\section*{Fission Reaction Rates}
  \newpage
  \renewcommand{\arraystretch}{0.5}
  \renewcommand\tabcolsep{1pt}
  \scriptsize
  \begin{center}

  \begin{longtable*}{@{}>{\small}l|>{\small}c>{\small}c>{\small}c>{\small}c>{\small}c>{\small}c@{}}
  \caption{Fission Reaction Rates from ENDF/B-VII.1, JEFF-3.1.2, JENDL-4.0, ROSFOND-2010, CENDL-3.1 and  EAF-2010  \cite{11Chad,11Kon,11Shi,07Zab,11Zhi,10Sub}  at {\it kT}=30 keV.} \label{RRTable18}
  \\
  \hline \hline \\
  \textbf{Material} & \textbf{ENDF/B-VII.1} & \textbf{JEFF-3.1.2} & \textbf{JENDL-4.0} & \textbf{ROSFOND-2010} & \textbf{CENDL-3.1}  & \textbf{EAF-2010} \\
  	& (cm$^{3}$/mole s) & (cm$^{3}$/mole s) & (cm$^{3}$/mole s) & (cm$^{3}$/mole s) & (cm$^{3}$/mole s) & (cm$^{3}$/mole s) 	\\
  \hline \\
  \endfirsthead
  \caption[]{Fission Reaction Rates (continued).}
  \\
  \hline \hline \\
  \textbf{Material} & \textbf{ENDF/B-VII.1} & \textbf{JEFF-3.1.2} & \textbf{JENDL-4.0} &  \textbf{ROSFOND-2010} & \textbf{CENDL-3.1}  & \textbf{EAF-2010} \\
  	& (cm$^{3}$/mole s) & (cm$^{3}$/mole s) & (cm$^{3}$/mole s) & (cm$^{3}$/mole s) & (cm$^{3}$/mole s) & (cm$^{3}$/mole s) 	\\
  \hline \\
\endhead
 86-Rn-222 &  &  &  & 5.061E+0 &  & 4.623E+4 \\
 88-Ra-223 & 9.087E+6 & 9.087E+6 & 9.087E+6 & 9.087E+6 &  & 9.087E+6 \\
 88-Ra-224 &  &  &  &  &  & 1.477E+3 \\
 88-Ra-225 &  &  &  &  &  & 8.208E+6 \\
 88-Ra-226 & 5.710E+1 & 5.710E+1 & 5.710E+1 & 5.710E+1 &  & 5.712E+1 \\
 88-Ra-228 &  &  &  & 1.641E+6 &  & 1.641E+6 \\
 89-Ac-225 & 1.225E+4$\pm$2.861E+4 &  & 1.225E+4 & 1.478E+3 &  & 1.477E+3 \\
 89-Ac-226 & 2.858E+6$\pm$6.384E+6 &  & 2.858E+6 & 8.994E+7 &  & 1.677E+7 \\
\hline \hline
 \end{longtable*}

 \end{center}
\normalsize

\section*{Capture Reaction Rates}
  \newpage
  \renewcommand{\arraystretch}{0.5}
  \renewcommand\tabcolsep{1pt}
  \scriptsize
  \begin{center}

  \begin{longtable*}{@{}>{\small}l|>{\small}c>{\small}c>{\small}c>{\small}c>{\small}c>{\small}c>{\small}c@{}}
  \caption{Capture Reaction Rates from ENDF/B-VII.1, JEFF-3.1.2, JENDL-4.0, ROSFOND-2010, CENDL-3.1,    EAF-2010  and  KADoNiS \cite{11Chad,11Kon,11Shi,07Zab,11Zhi,10Sub,06Dil}  at {\it kT}=30 keV.} \label{RRTable102}
  \\
  \hline \hline \\
  \textbf{Material} & \textbf{ENDF/B-VII.1} & \textbf{JEFF-3.1.2} & \textbf{JENDL-4.0} & \textbf{ROSFOND-2010} & \textbf{CENDL-3.1}  & \textbf{EAF-2010}  & \textbf{KADoNiS} \\
  	& (cm$^{3}$/mole s) & (cm$^{3}$/mole s) & (cm$^{3}$/mole s) & (cm$^{3}$/mole s) & (cm$^{3}$/mole s) & (cm$^{3}$/mole s) & (cm$^{3}$/mole s)	\\
  \hline \\
  \endfirsthead
  \caption[]{Capture Reaction Rates (continued).}
  \\
  \hline \hline \\
  \textbf{Material} & \textbf{ENDF/B-VII.1} & \textbf{JEFF-3.1.2} & \textbf{JENDL-4.0} &  \textbf{ROSFOND-2010} & \textbf{CENDL-3.1}  & \textbf{EAF-2010}  & \textbf{KADoNiS} \\
  	& (cm$^{3}$/mole s) & (cm$^{3}$/mole s) & (cm$^{3}$/mole s) & (cm$^{3}$/mole s) & (cm$^{3}$/mole s) & (cm$^{3}$/mole s) & (cm$^{3}$/mole s)	\\
  \hline \\
\endhead
 1-H -  1 & 3.112E+4$\pm$1.195E+3 & 3.109E+4 & 3.112E+4 & 3.112E+4 & 3.214E+4 & 3.108E+4 & 5.200E+4 \\
 1-H -  2 & 3.531E+2$\pm$2.236E+1 & 3.531E+2 & 4.173E+2 & 3.532E+2 & 3.528E+2 & 3.533E+2 & 5.320E+2 \\
 1-H -  3 &  &  &  &  &  &  &  \\
 2-He-  3 & 3.543E+3 & 4.070E+0 & 2.733E+3 & 3.860E+3 & 2.733E+3 & 2.734E+3 & 1.270E+3 \\
 2-He-  4 &  &  &  &  &  &  &  \\
 3-Li-  6 & 5.107E+3$\pm$4.958E+2 & 5.107E+3 & 5.197E+3 & 5.106E+3 &  & 5.196E+3 &  \\
 3-Li-  7 & 7.168E+3 & 7.168E+3 & 6.021E+3 & 7.167E+3 & 8.245E+3 & 7.167E+3 & 6.500E+3 \\
 4-Be-  7 &  &  &  &  &  & 4.564E+3 &  \\
\hline \hline
 \end{longtable*}

 \end{center}
\normalsize

\section*{ENDF/B-VII.1 Neutron Elastic Scattering  Values and Low-Fidelity Project Uncertainties}
  \newpage
  \renewcommand{\arraystretch}{0.5}
  \renewcommand\tabcolsep{1pt}
  \scriptsize
  \begin{center}

  \begin{longtable*}{@{}>{\small}l|>{\small}c>{\small}c>{\small}c>{\small}c>{\small}c@{}}
  \caption{ENDF/B-VII.1 Neutron Elastic Scattering Values and Low-Fidelity Project Uncertainties \cite{11Chad,08Lit}.} \label{LFTable2}
  \\
  \hline \hline \\
  \textbf{Material} & \textbf{$\sigma^{therm}$} & \textbf{RI} & \textbf{$\sigma^{Maxw}$(30 keV)} & \textbf{RR(30 keV)} & \textbf{$\sigma^{Maxw}$(1420 keV)}   \\
  	& (barns) & (barns) & (barns) & (cm$^{3}$/mole s) & (barns)	\\
  \hline \\
  \endfirsthead
  \caption[]{ENDF/B-VII.1 Neutron Elastic Scattering  Values (continued).}
  \\
  \hline \hline \\
  \textbf{Material} &  \textbf{$\sigma^{therm}$} & \textbf{RI} & \textbf{$\sigma^{Maxw}$(30 keV)} & \textbf{RR(30 keV)} & \textbf{$\sigma^{Maxw}$(1420 keV)}   \\
  	& (barns) & (barns) & (barns) & (cm$^{3}$/mole s) & (barns)  	\\
  \hline \\
\endhead
 1-H -  1 & 3.008E+1$\pm$8.911E-2 & 2.645E+2$\pm$9.234E-1 & 1.454E+1$\pm$4.856E-2 & 2.967E+9$\pm$9.911E+6 & 4.560E+8$\pm$6.570E+6 \\
 1-H -  2 & 4.235E+0$\pm$8.471E-2 & 5.342E+1$\pm$9.711E-1 & 3.663E+0$\pm$4.750E-2 & 6.474E+8$\pm$8.395E+6 & 3.908E+8$\pm$5.787E+6 \\
 1-H -  3 & 1.943E+0 & 2.933E+1 & 1.858E+0 & 3.096E+8 & 3.096E+8 \\
 2-He-  3 & 3.862E+0 & 4.871E+1 & 2.888E+0 & 4.813E+8 & 4.813E+8 \\
 2-He-  4 & 8.635E-1 & 2.216E+1 & 8.751E-1 & 1.413E+8 & 1.413E+8 \\
 3-Li-  6 & 7.777E-1$\pm$1.555E-2 & 1.782E+1$\pm$3.364E-1 & 1.059E+0$\pm$1.840E-2 & 1.650E+8$\pm$2.868E+6 & 2.233E+8$\pm$6.286E+6 \\
 3-Li-  7 & 1.040E+0 & 2.135E+1 & 1.221E+0 & 1.884E+8 & 1.884E+8 \\
 4-Be-  7 & 2.042E+1$\pm$6.126E-1 & 2.054E+2$\pm$6.071E+0 & 8.379E+0$\pm$2.395E-1 & 1.293E+9$\pm$3.695E+7 & 2.947E+8$\pm$1.214E+7 \\
\hline \hline
 \end{longtable*}

 \end{center}
\normalsize

\section*{ENDF/B-VII.1 Neutron Fission  Values and Low-Fidelity Project Uncertainties}
  \newpage
  \renewcommand{\arraystretch}{0.5}
  \renewcommand\tabcolsep{1pt}
  \scriptsize
  \begin{center}

  \begin{longtable*}{@{}>{\small}l|>{\small}c>{\small}c>{\small}c>{\small}c>{\small}c@{}}
  \caption{ENDF/B-VII.1 Neutron Fission  Values and Low-Fidelity Project Uncertainties \cite{11Chad,08Lit}.} \label{LFTable18}
  \\
  \hline \hline \\
  \textbf{Material} & \textbf{$\sigma^{therm}$} & \textbf{RI} & \textbf{$\sigma^{Maxw}$(30 keV)} & \textbf{RR(30 keV)} & \textbf{$\sigma^{Maxw}$(1420 keV)}   \\
  	& (barns) & (barns) & (barns) & (cm$^{3}$/mole s) & (barns)	\\
  \hline \\
  \endfirsthead
  \caption[]{ENDF/B-VII.1 Neutron Fission   Values (continued).}
  \\
  \hline \hline \\
  \textbf{Material} &  \textbf{$\sigma^{therm}$} & \textbf{RI} & \textbf{$\sigma^{Maxw}$(30 keV)} & \textbf{RR(30 keV)} & \textbf{$\sigma^{Maxw}$(1420 keV)}   \\
  	& (barns) & (barns) & (barns) & (cm$^{3}$/mole s) & (barns)  	\\
  \hline \\
\endhead

 88-Ra-223 & 7.003E-1 & 1.062E+0 & 6.284E-2 & 9.087E+6 & 6.284E-2 \\
 88-Ra-224 &  &  &  &  &  \\
 88-Ra-225 &  &  &  &  &  \\
 88-Ra-226 & 7.002E-6 & 1.187E-2 & 3.949E-7 & 5.710E+1 & 3.949E-7 \\
 89-Ac-225 & 5.003E-4 & 2.159E-1 & 8.471E-5 & 1.225E+4 & 8.471E-5 \\
 89-Ac-226 & 2.001E+0 & 8.705E+0 & 1.977E-2 & 2.858E+6 & 1.977E-2 \\
 89-Ac-227 & 1.000E-4$\pm$6.668E-5 & 1.167E-1$\pm$1.831E-2 & 6.786E-6$\pm$9.368E-11 & 9.812E+2$\pm$1.354E-2 & 2.111E-2$\pm$3.897E-3 \\
 90-Th-227 & 2.021E+2$\pm$1.301E+1 & 3.310E+2$\pm$8.590E-1 & 6.006E-1$\pm$1.716E-1 & 8.685E+7$\pm$2.481E+7 & 3.864E-1$\pm$6.490E-2 \\
\hline \hline
 \end{longtable*}

 \end{center}
\normalsize

\section*{ENDF/B-VII.1 Neutron Capture  Values and Low-Fidelity Project Uncertainties}
  \newpage
  \renewcommand{\arraystretch}{0.5}
  \renewcommand\tabcolsep{1pt}
  \scriptsize
  \begin{center}

  \begin{longtable*}{@{}>{\small}l|>{\small}c>{\small}c>{\small}c>{\small}c>{\small}c@{}}
  \caption{ENDF/B-VII.1 Neutron Capture  Values and Low-Fidelity Project Uncertainties \cite{11Chad,08Lit}.} \label{LFTable102}
  \\
  \hline \hline \\
  \textbf{Material} & \textbf{$\sigma^{therm}$} & \textbf{RI} & \textbf{$\sigma^{Maxw}$(30 keV)} & \textbf{RR(30 keV)} & \textbf{$\sigma^{Maxw}$(1420 keV)}   \\
  	& (barns) & (barns) & (barns) & (cm$^{3}$/mole s) & (barns)	\\
  \hline \\
  \endfirsthead
  \caption[]{ENDF/B-VII.1 Neutron Capture  Values (continued).}
  \\
  \hline \hline \\
  \textbf{Material} &  \textbf{$\sigma^{therm}$} & \textbf{RI} & \textbf{$\sigma^{Maxw}$(30 keV)} & \textbf{RR(30 keV)} & \textbf{$\sigma^{Maxw}$(1420 keV)}   \\
  	& (barns) & (barns) & (barns) & (cm$^{3}$/mole s) & (barns)  	\\
  \hline \\
\endhead

 1-H -  1 & 3.320E-1$\pm$8.475E-3 & 1.492E-1$\pm$3.812E-3 & 1.525E-4$\pm$5.248E-6 & 3.112E+4$\pm$1.071E+3 & 3.892E-5$\pm$7.842E-6 \\
 1-H -  2 & 5.057E-4$\pm$1.517E-5 & 2.640E-4$\pm$2.192E-5 & 1.998E-6$\pm$1.023E-7 & 3.531E+2$\pm$1.808E+1 & 1.025E-5$\pm$6.351E-6 \\
 1-H -  3 &  &  &  &  &  \\
 2-He-  3 & 5.499E-5$\pm$2.749E-5 & 3.637E-4$\pm$1.790E-4 & 2.126E-5$\pm$1.037E-5 & 3.543E+3$\pm$1.728E+3 & 8.244E-5$\pm$4.122E-5 \\
 2-He-  4 &  &  &  &  &  \\
 3-Li-  6 & 3.850E-2$\pm$3.080E-3 & 1.736E-2$\pm$1.385E-3 & 3.276E-5$\pm$2.346E-6 & 5.107E+3$\pm$3.658E+2 & 1.253E-5$\pm$9.879E-6 \\
 3-Li-  7 & 4.540E-2 & 2.045E-2 & 4.645E-5 & 7.168E+3 & 4.645E-5 \\
 4-Be-  7 &  &  &  &  &  \\
\hline \hline
 \end{longtable*}

 \end{center}
\normalsize

\end{document}